\documentclass[conference]{IEEEtran}

\usepackage{epsfig,endnotes}

\usepackage[ruled, vlined, linesnumbered]{algorithm2e}
\usepackage{setspace}
\usepackage{graphicx}
\usepackage{mathptmx}
\usepackage{amsmath}
\usepackage{amssymb}
\usepackage{subfigure}
\usepackage{array}
\usepackage{multirow}
\usepackage{fontawesome}
\usepackage{makecell}
\usepackage{amsfonts}
\usepackage{grffile}
\usepackage{float} 
\usepackage{chapterbib}
\usepackage{verbatim}
\usepackage{docmute}
\usepackage{bm}
\usepackage{xcolor,colortbl}
\usepackage{threeparttable}
\usepackage{booktabs}
\usepackage{enumitem}
\usepackage{colortbl}
\usepackage{hyperref}
\usepackage{caption}
\usepackage[T1]{fontenc}
\setenumerate[1]{itemsep=0pt,partopsep=0pt,parsep=\parskip,topsep=5pt}
\setitemize[1]{itemsep=0pt,partopsep=0pt,parsep=\parskip,topsep=5pt}
\setdescription{itemsep=0pt,partopsep=0pt,parsep=\parskip,topsep=5pt}
\newcommand{\tool}[1]{{\textsc{#1}}\xspace}
\newcommand{\verifi}{\tool{VeriFi}}
\newcommand{\lab}[1]{\textquotesingle{#1}\textquotesingle}

\usepackage{tcolorbox}
\tcbuselibrary{most}

\usepackage{amsmath,amsfonts,bm}

\def\eqref#1{equation~(\ref{#1})}
\def\Eqref#1{Eq.~(\ref{#1})}

\def\1{\bm{1}}

\def\vr{{\bm{r}}}

\def\vw{{\bm{w}}}
\def\vx{{\bm{x}}}

\def\vz{{\bm{z}}}

\DeclareMathAlphabet{\mathsfit}{\encodingdefault}{\sfdefault}{m}{sl}
\SetMathAlphabet{\mathsfit}{bold}{\encodingdefault}{\sfdefault}{bx}{n}

\def\gD{{\mathcal{D}}}

\def\sD{{\mathbb{D}}}
\def\sE{{\mathbb{E}}}

\newcommand{\etens}[1]{\mathsfit{#1}}

\def\etC{{\etens{C}}}

\definecolor{xsgray}{RGB}{190,190,190}
\newcommand{\commentout}[1]{}
\usepackage[normalem]{ulem}
\usepackage{soul}

\usepackage{marginnote}

\begin{document}

\date{}

\title{\verifi: Towards Verifiable Federated Unlearning}


\author{\IEEEauthorblockN{Xiangshan Gao}
\IEEEauthorblockA{Zhejiang University\\
corazju@zju.edu.cn}
\and
\IEEEauthorblockN{Xingjun Ma}
\IEEEauthorblockA{Fudan University\\
xingjunma@fudan.edu.cn}
\and
\IEEEauthorblockN{Jingyi Wang}
\IEEEauthorblockA{Zhejiang University\\
wangjyee@zju.edu.cn}
\and
\IEEEauthorblockN{Youcheng Sun}
\IEEEauthorblockA{University of Manchester\\
youcheng.sun@manchester.ac.uk}
\and
\IEEEauthorblockN{Bo Li}
\IEEEauthorblockA{UIUC\\
lbo@illinois.edu} 
\and
\IEEEauthorblockN{Shouling Ji}
\IEEEauthorblockA{Zhejiang University\\
sji@zju.edu.cn}
\and
\IEEEauthorblockN{Peng Cheng}
\IEEEauthorblockA{Zhejiang University\\
lunarheart@zju.edu.cn}
\and
\IEEEauthorblockN{Jiming Chen}
\IEEEauthorblockA{Zhejiang University\\
cjm@zju.edu.cn}
}

\maketitle
\pagestyle{plain}

\begin{abstract}
Federated learning (FL) has emerged as a privacy-aware collaborative learning paradigm where participants jointly train a powerful model without sharing their private data. One desirable property for FL is the implementation of the \emph{right to be forgotten (RTBF)}, i.e., a leaving participant has the right to request to delete its private data from the global model. 
Recently, several server-side unlearning methods have been proposed to remove a leaving participant's gradients from the global model. 
However, \emph{unlearning itself may not be enough to implement RTBF unless the unlearning effect can be independently verified}, an important aspect that has been overlooked in the current literature.
In this paper, we prompt the concept of \emph{verifiable federated unlearning}, and propose \verifi, a unified framework integrating federated unlearning and verification that allows systematic analysis of the unlearning and quantification of its effect, with different combinations of multiple unlearning and verification methods. In \verifi, the leaving participant is granted the \emph{right to verify (RTV)}, that is, the participant notifies the server before leaving, then actively verifies the unlearning effect in the next few communication rounds. The unlearning is done at the server side immediately after receiving the leaving notification, while the verification is done locally by the leaving participant via two steps: \emph{marking} and \emph{checking}. The marking step injects carefully-designed \emph{markers} to fingerprint the leaving participant's data, while the checking step examines the change of the global model's performance on the markers.

Based on \verifi, we conduct the first systematic and large-scale study for verifiable federated unlearning, considering 7 unlearning methods and 5 verification methods 
that cover existing, adapted and newly proposed ones for both unlearning and verification.
Particularly, the newly proposed methods include a more efficient and FL-friendly unlearning method $^u$S2U, and two more effective and robust non-invasive-verification methods $^v$FM and $^v$EM (without training controllability or external data, without white-box model access or introducing security hazard).
We extensively evaluate \verifi on 7 datasets, including both (natural/facial/medical) images and audios, and 4 types of deep learning models, including both Convolutional Neural Networks (CNNs) and Recurrent Neural Networks (RNNs). 
Our analysis establishes important empirical understandings and evidence for more trustworthy federated unlearning.
\end{abstract}

\section{Introduction}\label{sec:intro}

Federated learning (FL) is a collaborative learning paradigm that allows participants to train a powerful machine learning model jointly without sharing their private data \cite{bonawitz2019towards, kairouz2019advances,yang2019federated}.
This privacy-preserving nature of FL makes it an ideal choice for real-world privacy-sensitive collaborations in finance \cite{WeBank}, healthcare \cite{xu2021federated, brisimi2018federated}, insurance \cite{wang2019interpret} and many other fields. One essential requirement of FL is the participants' \emph{``right to be forgotten'' (RTBF)}, which has been stated explicitly in the European Union General Data Protection Regulation (GDPR) \cite{GDPR, regulation2018general} and the California Consumer Privacy Act (CCPA) \cite{harding2019understanding}. That is, a participant has the right to request a deletion of its private data.
Arguably, one may worry that its private data will be memorized by the global model and continue to be exploited even after leaving the federation. 
As leaving/joining is a common behavior in FL, it is thus necessary to ensure that every participant can \textbf{join and leave the federation freely, and more importantly, with no concerns}. However, so far, participants have difficulty exercising the RTBF in existing FL frameworks, which might discourage potential participants to join the federation.

The concept of \emph{machine unlearning} \cite{bourtoule2021machine,shintre2019making} has recently been proposed to remove data from a machine learning model. 
Several unlearning methods are designed to actively unlearn certain data from a trained model.
A simple yet costly approach for unlearning is to retrain the model from scratch with the requested data being removed from the training set \cite{bourtoule2021machine}. It can be made more efficient if the model is trained on summarized (e.g., aggregates of summations) or partitioned subsets rather than individual training samples, in which case, the model only needs to be updated on the subset(s) associated with the requested samples \cite{cao2015towards, ginart2019making}. The above methods are less practical for large-scale datasets, although advanced data partitioning or intermediate model breakpoint strategies may help \cite{bourtoule2021machine, he2021deepobliviate}. 
More recently, machine unlearning has been extended to the FL setting, a.k.a., \emph{federated unlearning} \cite{liu2021federaser}, which is arguably more challenging. In FL, 1) the global model is updated based on the aggregated rather than the raw gradients; 2) FL can have a large number of participants; and 3) different participants may have similar, or to some extent, shared training samples.
Consequently, simple gradient-based methods such as subtracting the reconstructed or dummy gradients of the leaving participant may harm the original task or introduce new privacy threats into FL \cite{liu2021federaser, liu2020learn}.

Moreover, federated unlearning is only one side of the coin for the RTBF. A more concerning question from a participant's perspective is: \textbf{how to make sure that my data has indeed been forgotten, hopefully in a verifiable and measurable way,} which we believe is the core of establishing mutual trust in FL.
Unfortunately, this important aspect has been largely overlooked in the FL literature. 
In traditional machine unlearning, the unlearning effect can be simply verified by the model's performance (e.g., accuracy and loss) on the unlearned data or additionally injected backdoor data \cite{bourtoule2021machine, cao2015towards, sommer2020towards}. 
However, in FL, the accuracy and loss may hardly change when only one or a few participants left the federation, owing to the contribution of other participants. Besides, it is not secure in FL to use backdoor solely for the purpose of unlearning verification as it might introduce new security threats into the commonly contributed and shared global model (see Appendix \ref{Security risk of $^v$BN}).
So far, it still lacks understanding of how to \emph{effectively} and \emph{reliably}  verify that the data has indeed been deleted after unlearning. In fact, due to the lack of a unified, holistic and all-round FL verification framework, several key fundamental questions for trustworthy RTBF in FL remain unexplored:

\begin{itemize}
\label{rqs}
    \item \emph{Federated Unlearning.} Is federated unlearning necessary or might natural forgetting be enough to forget the leaving participant's data?
    \item \emph{Unlearning Verification.} Do we need more sophisticated methods 
    or simple methods like checking the global model's performance on the leaving participant's data are enough to measure and verify the unlearning effect?
    \item \emph{Practical Choice.} What are the most effective combination(s) of unlearning and verification methods that can effectively unlearn, clearly verify, while causing minimal negative impact on the original task?
\end{itemize} 

\begin{figure*}[!htb]
\centering
\includegraphics[width=2.0\columnwidth]{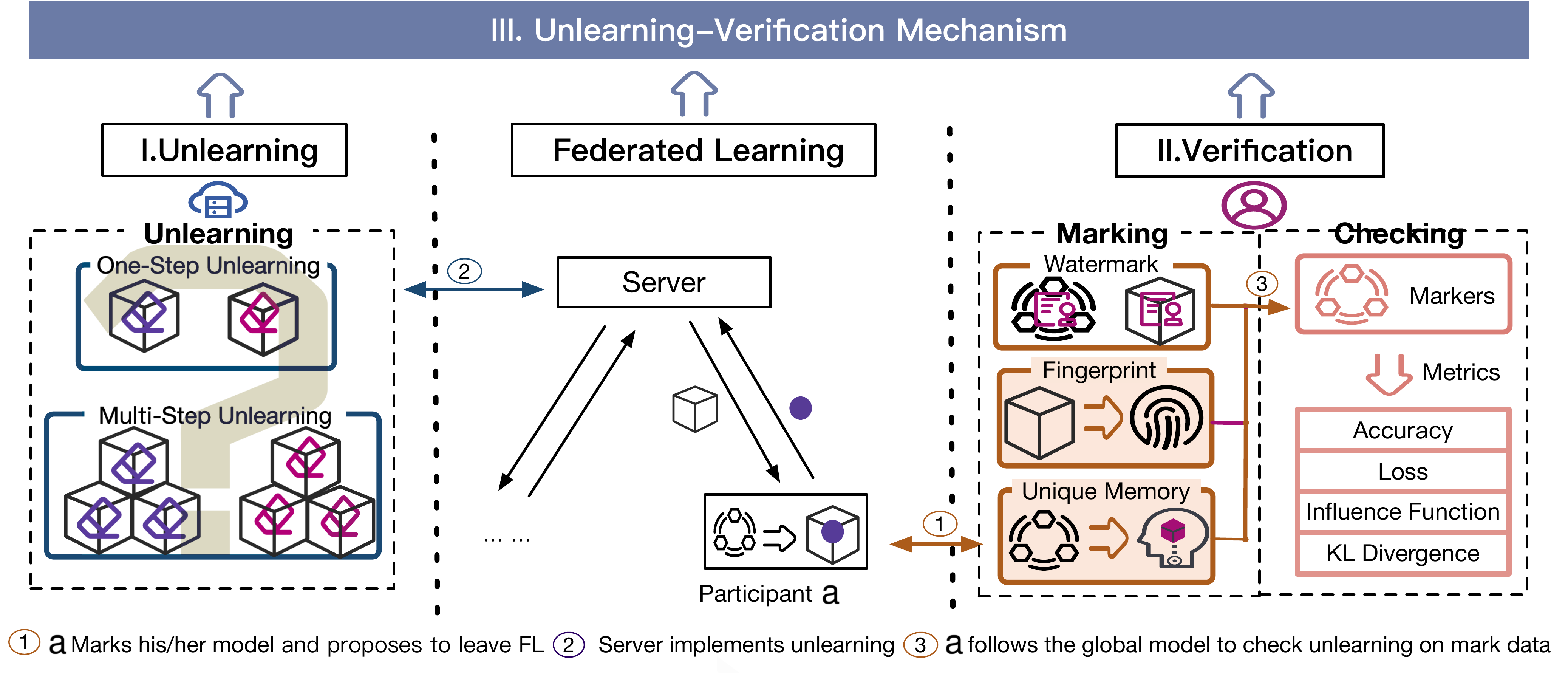}
\caption{Overview of the proposed \verifi framework and its three key modules: 1) unlearning module; 2) verification module; and 3)   unlearning-verification mechanism. The standard ``Federated Learning'' procedure is further illustrated in Fig. \ref{withdraw_federated_learning}.}
\label{framework_down}
\end{figure*}

To answer the above questions, in this paper, we promote the concept of \emph{verifiable federated unlearning}, which treats verification as important as unlearning and grants the participant the \emph{``right to verify'' (RTV)}.
Specifically, we design and implement \verifi, a unified framework for verifiable federated unlearning. The core of \verifi contains 1) a federated unlearning module; 2) a verification module with two key verification steps, namely \emph{marking} and \emph{checking}; and 3) a generic \emph{unlearning-verification} mechanism applicable to common FL frameworks. 
Fig.~\ref{framework_down} provides an overview of \verifi.
The unlearning module can be any  unlearning\footnote{Without ambiguity, we use ``unlearning'' instead of ``federated unlearning'' for simplicity.} method adopted at the server size that erases the information of the leaving participant's data (which we call ``leaving data''). 
The marking step of the verification module injects/tags specifically selected or designed patterns or training examples as \emph{markers}. 
The checking step of the verification module then verifies the degree of unlearning based on different verification metrics defined w.r.t. the global model and the markers. The \emph{unlearning-verification} mechanism integrates all the above steps into a chained pipeline and specifies when and what to mark, and who and when to unlearn and verify.

With \verifi, we bring together a comprehensive set of unlearning and verification methods, including not only existing ones but also many adapted from other fields, as well as newly proposed in this paper. We conduct the first systematic study on the practicality of different combinations of unlearning and verification methods for verifiable federated unlearning.
\textbf{For unlearning}, we study the limitations of existing one-step (e.g., differential privacy\footnote{ Although $^u$DP cannot ensure completely zero memory in machine unlearning \cite{bourtoule2021machine}, we still explored its practical unlearning effect in federated unlearning for the purpose of completeness.}) and multi-step (e.g., retraining and gradient subtraction) unlearning methods, such as high cost and significant negative impact on the original task. We also propose a more efficient and FL-friendly one-step unlearning method \emph{scale-to-unlearn} ($^u$S2U)\footnote{We use the $^u$ symbol to indicate unlearning methods.}.
$^u$S2U scales down the leaving participant's gradients/parameters to trigger the global model to erase its memorization of the participant. 
Verification consists of two steps: marking and checking.
\textbf{For marking}, the existing method leverages backdoored samples to verify the unlearning effect \cite{sommer2020towards}, which is unsuitable for FL as backdoor methods are invasive methods that could introduce global threats to all FL participants. 
We consider this backdoor-based verification method in \verifi as a comparison, and further propose two non-invasive unique memory-based methods. 
The two proposed verification methods verify the unlearning effect based on the sensitive performance of the global model on a specific subset of the leaving data. 
Moreover, we also adapt existing watermark and fingerprint methods proposed for deep learning intellectual property protection as verification methods for federated unlearning. 
We systematically analyze the pros and cons of these marking methods in \verifi. 
\textbf{For checking}, we explore loss, accuracy, influence function (IF) \cite{koh2017understanding} and Kullback–Leibler (KL) divergence \cite{goldberger2003efficient} to measure the performance change on the marked data (i.e., markers) before and after unlearning.
Our extensive evaluation and analyses provide answers to the three fundamental questions mentioned earlier, and establish the empirical foundation for verifiable and trustworthy federated unlearning.

In summary, our main contributions are:
\begin{itemize}[leftmargin=*, topsep =2 pt]
\item We design the first unlearning-verification framework \verifi for verifiable federated unlearning.
\verifi grants FL participants the \emph{right to verify}, i.e., the verification of the unlearning effect when leaving the federation. \verifi introduces a unified mechanism that allows quantitative measurement on the effectiveness of different combinations of unlearning and verification methods.
\item With \verifi, we identify the limitations of existing unlearning and verification methods, and propose a more efficient and FL-friendly unlearning method $^u$S2U and two more effective and robust non-invasive \emph{unique memory} based verification methods ($^v$EM and $^v$FM)\footnote{We use the $^v$ symbol to indicate verification methods.}. The advantages of the three proposed methods are also demonstrated by our extensive experiments.
\item With \verifi, we systemically study 7 unlearning methods and 5 verification methods (i.e., 5 marking methods and 4 checking metrics) with both Convolutional Neural Networks (CNNs) and Recurrent Neural Networks (RNNs) on 7 datasets, including 3 natural image, 1 facial image, 1 audio and 2 medical image datasets. Our extensive study unveils the necessity, potentials and limitations of different federated unlearning and verification methods.

\end{itemize}

\section{Preliminaries}\label{preliminary}
\newenvironment{shrinkeq}[1]
{ \bgroup
  \addtolength\abovedisplayshortskip{#1}
  \addtolength\abovedisplayskip{#1}
  \addtolength\belowdisplayshortskip{#1}
  \addtolength\belowdisplayskip{#1}}
{\egroup\ignorespacesafterend}
\subsection{Federated Learning}
\begin{figure}[!htb]
  \centering
  \includegraphics[width=1.0\columnwidth]{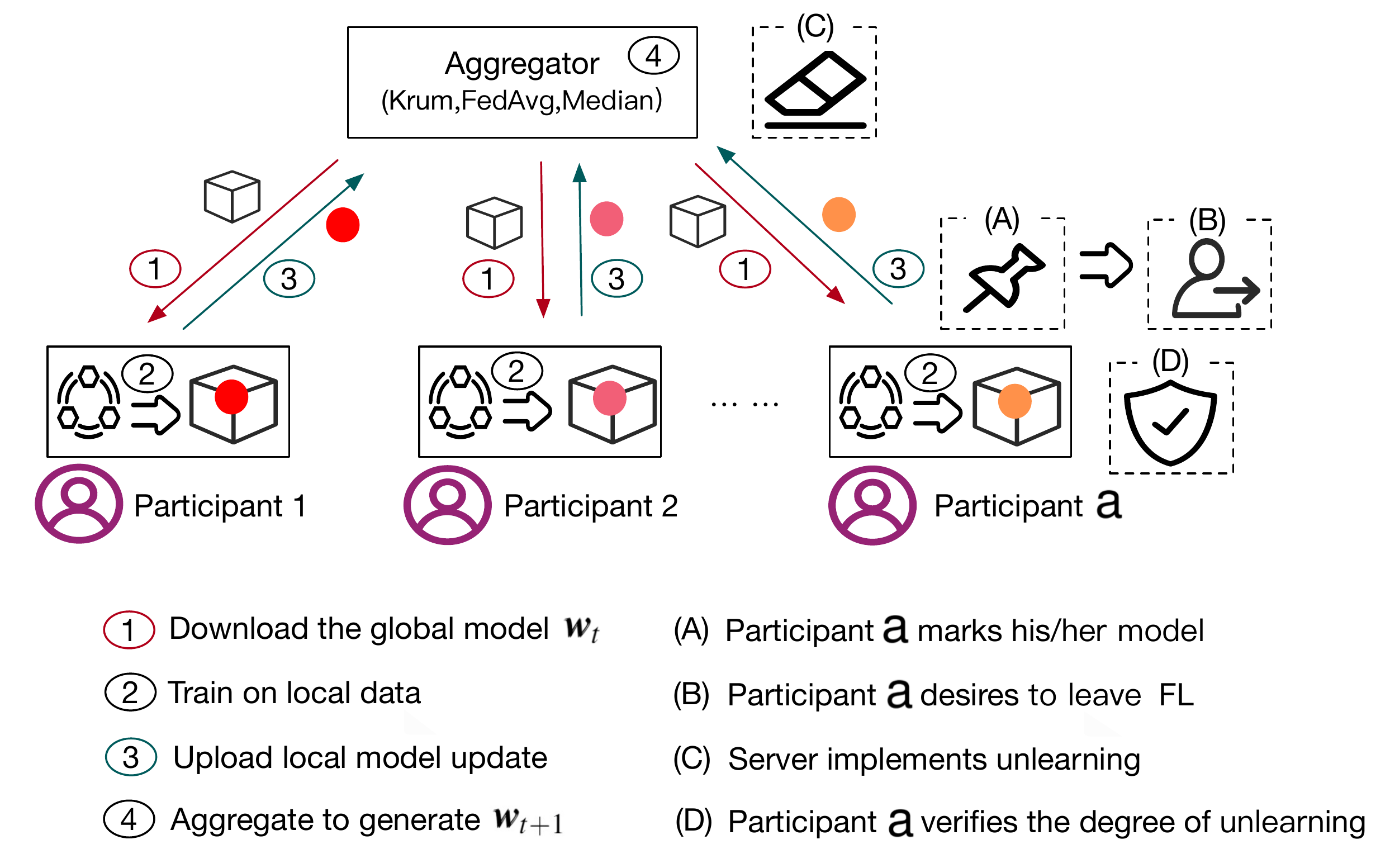}
  \caption{Collaboration and learning in federated learning.}
  \label{withdraw_federated_learning}
\end{figure}

In FL, a number of participants jointly train a global model by communicating gradients or model parameters with a central server. At each communication round, the participants download the global model from the server, perform a certain number of local updates on their private data, and then upload the accumulated local updates (gradients) to the server. The server then aggregates (e.g., using FedAvg \cite{mcmahan2017communication}) the accumulated local updates to update the global model. The complete FL procedure is illustrated in Fig. \ref{withdraw_federated_learning}. The participants' private data is protected during the entire FL process, as it never leaves the local devices.

Let $[n]=\{1,...,n\} $ be the set of $n$ participants with each participant owning a private dataset $D_{i}$ for $i \in [n]$, and $\gD = D_{1} \cup D_{2} \cup \cdots D_{n}$ is the full training dataset.
At the $t$-th communication round, the $i$-th participant first downloads the global model $\vw_t$, and then performs local update(s), e.g., using Stochastic Gradient Descent (SGD), on the local data $D_{i}$ to obtain an updated local model $\vw_{t+1}^{(i)}$. The accumulated gradient, $\vw_{t+1}^{(i)} - \vw_t$, is then sent to the server for the global model update, e.g., using FedAvg~\cite{mcmahan2017communication} as follows:
\begin{shrinkeq}{-1ex}
\begin{equation}\label{eq:fl}
\vw_{t+1} = \vw_t + \frac{1}{n}\sum_{i \in [n]}(\vw_{t+1}^{(i)} - \vw_t).
\end{equation}
\end{shrinkeq}
Besides FedAvg, other aggregation rules are proposed for Byzantine-robust FL: Krum \cite{blanchard2017machine}, Median \cite{yin2018byzantine}, Bulyan \cite{guerraoui2018hidden} and Trimmed Mean \cite{yin2018byzantine}. 
Meanwhile, FL can be either horizontal where the participants share the same feature space but own different data samples, or vertical where the participants share the same data sample IDs but possess different features.
In this work, we focus on a typical horizontal FL setting with FedAvg, as defined in \Eqref{eq:fl}.

\subsection{Federated Unlearning and Verification}\label{Unlearning verification --- Watermark}

\noindent\textbf{Federated Unlearning.} It has been shown that deep neural networks have both memorization and forgetting effects \cite{zhang2017understanding,arpit2017closer,kirkpatrick2017overcoming}, i.e., they naturally memorize information about the training data and so naturally forget the removed data (from the training dataset) during training.
Different from natural forgetting\footnote{The participant leaves with no active unlearning conducted by the server.}, machine unlearning explicitly forces a model to forget its memorization of a target (requested to delete) subset of training samples \cite{bourtoule2021machine}. Intuitively, unlearning can be achieved by (re)training the model on the updated dataset with the requested samples removed. In traditional machine learning, this can be done via expensive retraining, or more efficient partition/breakpoint based learning with data partitions/aggregates \cite{cao2015towards, ginart2019making, bourtoule2021machine,he2021deepobliviate}. Noise can also be used to smooth out the memorization of particular samples \cite{shintre2019making}. However, in FL, information is shared via gradients. This motivates the two pioneering works \cite{liu2021federaser, liu2020learn} in federated unlearning to subtract the calibrated or generated gradients of the leaving participant to unlearn information. We will incorporate and test the two methods (as well as the costly retraining) in \verifi and propose a more effective unlearning method for FL.
More detailed analysis about the pros and cons of the unlearning methods in \verifi can be found in Section \ref{unlearning} and \ref{evaluation}.

\noindent\textbf{Unlearning Verification.}
Intuitively, the effectiveness of unlearning can be verified by the change in the model's performance before and after unlearning. In existing works, loss or accuracy on the leaving data is often used to achieve the verification purpose \cite{bourtoule2021machine}. Unlearning can also be verified on a set of backdoored training samples \cite{sommer2020towards} obtained via a backdoor attack, which is essentially a data poisoning process that injects a trigger pattern into a small subset of training data so as to trick the model into memorizing the correlation between the pattern and a target class \cite{gu2017badnets,chen2017targeted}. Suppose the trigger pattern is $\vr$ and its associated backdoor target class is $y_{target}$. Once the trigger is learned by the model $f$, the model will constantly predict the target class on any samples attached with the trigger pattern:
\begin{shrinkeq}{-1ex}
\begin{equation}
\arg\max f(\vx \oplus \vr) = y_{target}, \; \forall (\vx,y) \in \gD,
\end{equation}
\end{shrinkeq}
where, the model $f$ outputs the class probabilities, the operation $\vx \oplus \vr$ produces a backdoored version of $\vx$, $(\vx,y)$ is an input-label pair, and $\gD$ is the training dataset in traditional machine learning. If the unlearning is effective, then the model will forget the backdoor correlation and predict the correct class instead:
\begin{shrinkeq}{-1ex}
\begin{equation}
\arg\max \overline{f}(\vx \oplus \vr) = y, \; \forall (\vx,y) \in \gD,
\end{equation}
\end{shrinkeq}
where $\overline{f}$ denotes the model obtained after unlearning and $y$ is the correct class of $\vx$.

Although several unlearning methods have been proposed, the challenge and potential issues of unlearning verification have not been thoroughly studied, especially in FL. In fact, \cite{sommer2020towards} is the only work that has investigated the verification problem, however, it was conducted in traditional machine unlearning. It proposes to use backdoored samples to obtain more sensitive verification. Considering the high security risk (could backdoor all participants) of backdoor techniques, it is thus
not ideal to use backdoor verification in FL.

We also adapt and study two plausible concepts from the deep learning intellectual property (IP) protection domain for unlearning verification: watermarking \cite{uchida2017embedding, zhang2018protecting} and fingerprinting \cite{cao2019ipguard}.
Watermarking is an invasive technique that embeds owner-specific binary string or backdoor triggers into the model parameters to help determine the ownership of the model at a later (post-deployment) stage, while fingerprinting generates new samples to fingerprint the model's unique properties like decision boundary  \cite{cao2019ipguard}.
In this work, we specially design and adapt these two types of techniques for federated unlearning verification.
More systematic analysis of different verification methods can be found in Section \ref{verification} and \ref{sec:markers_experiment}.
\section{Proposed \verifi Framework}\label{Framework}

\begin{figure*}[t]
\centering
\includegraphics[width=0.94\textwidth]{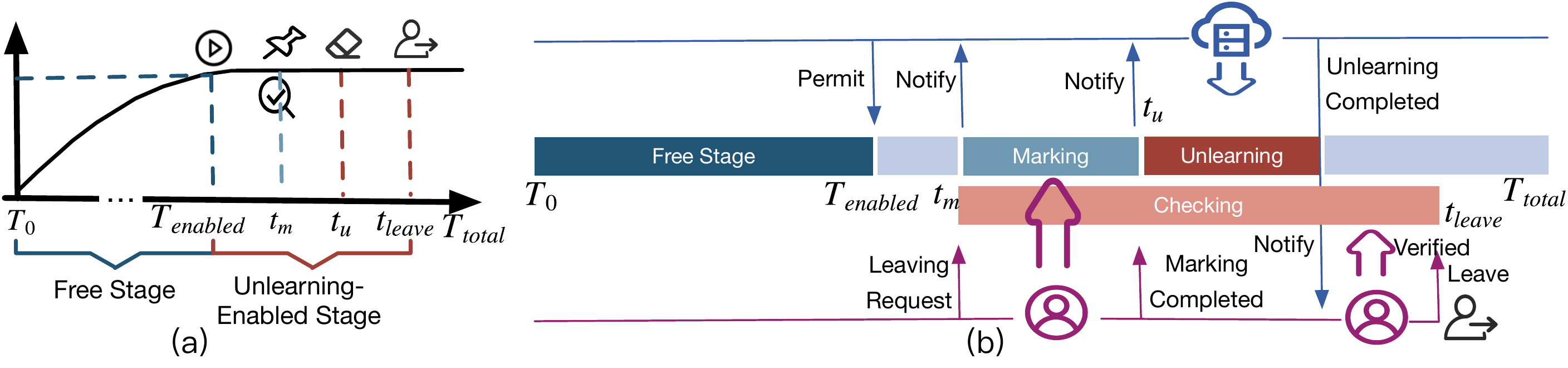}
\caption{The proposed unlearning-verification mechanism.}
\label{timeline}
\end{figure*}

In this section, we present our \verifi framework in detail. Lying at the core of \verifi is our proposed \emph{unlearning-verification mechanism}. As illustrated in Fig. \ref{timeline}, the mechanism defines the timeline when unlearning and verification should be performed, and by whom, i.e., the central server or the leaving participant (``leaver''). Here, we focus on unlearning the leaver and his/her verification in FL in Fig. \ref{timeline}.

\subsection{Unlearning-Verification Mechanism}
Suppose the entire FL process consists of $T_{total}$ communication rounds. As shown in Fig. \ref{timeline}(a), the mechanism divides the entire process into two stages, including a free stage ($[T_{0}, T_{enabled})$) and an unlearning-enabled stage ($[T_{enabled}, T_{total}]$).
The free stage refers to an early FL stage where the global model has not yet converged to a good solution. In this stage, all participants can join and leave the federation freely without activating the unlearning mechanism, as in this stage, the next round of training often overwrites the model's memorization at the previous rounds. Leaving the federation after $T_{enabled}$ will activate the unlearning and verification process, as at this time, the model's memorization of the private data is stabilized. Note that joining the federation at this stage should also be carefully examined as it is a \emph{harvest stage} where small contribution can receive a big reward, i.e., a high-performance global model. Here, we only focus on leaving and unlearning.

Fig. \ref{timeline}(b) shows the pipelined unlearning-verification mechanism with a single leaving participant\footnote{\verifi is easily extended to the situation where multiple leavers require to be forgotten and verified.} . The detailed steps can be found in Mechanism \ref{alg:UVP}.
In this paper, we focus on one leaving participant per round while leaving more complex scenarios to future work.
Specifically, the leaving participant (denoted by $\mathsf{a}$) first notifies the server about the leaving at $t_m$ (Step 2). 
Meanwhile, the leaving participant applies  a marking method to mark the data (e.g., private training samples, triggers or model parameters) that needs to be checked against unlearning (Step 3.a). We call the marked data \emph{`markers'}.
Once the marked model is uploaded to the server (Step 3.b), the leaving participant notifies the server to apply the unlearning method to unlearn its data (Step 3.c). Note that the server may or may not be aware of the existence of markers since the verification right is in the hands of the participants, not the server. The server-side unlearning may last for more than one communication round (Step 4.a).
The participant will actively check the unlearning effect on the markers immediately after marking is completed (Step 4.b).
After a few rounds of checking at $t_{leave}$, the participant will leave with assured privacy (Step 5.a) or distrust (Step 5.b), depending on whether the expected unlearning effect on the marker is satisfied.

\begin{algorithm}[!htb]
    \SetAlgoNoLine
    \setstretch{0.6}
    \DontPrintSemicolon
    \renewcommand{\algorithmcfname}{Mechanism}
    \caption{Federated Unlearning-Verification}
    \label{alg:UVP}
    \KwIn{Unlearning-enabled stage $[T_{enabled},T_{total})$, marking starting point $t_m \in [T_{enabled}, T_{total})$, unlearning starting point $t_u  \in (t_m, t_{leave})$, leaving point $t_{leave} \in (t_u, T_{total})$, checking metric threshold $\delta$,
    marking function $\phi(\cdot)$,
    unlearning function $\varphi(\cdot)$,
    checking function $\psi(\cdot)$,
    aggregation rule $Agg(\cdot)$}
    \begin{enumerate}
    \item \textbf{Free stage:} Vanilla FL before $T_{enabled}$ 
    \item At $t_m$, participant $\mathsf{a}$ notifies the server to leave
    \item \textbf{Marking at $t_m$:}
    \begin{enumerate}
    \item $\mathsf{a}$ marks its local model $\widetilde{  \vw}_{t+1}^{(\mathsf{a})}$ $\leftarrow$  $\phi(\vw_{t+1}^{(\mathsf{a})})$
    \item $\mathsf{a}$ uploads the marking update  $ \widetilde{\vw}_{t+1}^{(\mathsf{a})} - \vw_t$ to server
    \item $\mathsf{a}$ notifies the server the completion of marking
    \end{enumerate}
    \item \textbf{Unlearning at $t_u \in [t_m+1, t_{leave})$:}
    \begin{enumerate}
    \item Server performs aggregation and unlearning:\\ $  \vw_{t+1}$ $\leftarrow$ $Agg \left(\varphi \left( \left \{\vw_{t+1}^{(i)}  - \vw_t  \right\}_{i=1}^n \right) \right)$
    \item \textbf{Checking}: $\mathsf{a}$ checks if unlearning is sufficient $  \psi(\vw_{t_m})  - \psi(\vw_{t+1}) \geq \delta$
    \end{enumerate}
    \item \textbf{Leaving at $t_{leave}$:}
    \begin{enumerate}
    \item $\mathsf{a}$ leaves with assured privacy if $  \psi(\vw_{t_m})  - \psi(\vw_{t+1}) \geq \delta$
    \item $\mathsf{a}$ leaves with distrust  if $  \psi(\vw_{t_m})  - \psi(\vw_{t+1}) < \delta$
    \end{enumerate}
    \end{enumerate}
\end{algorithm}
\setlength{\textfloatsep}{0.1cm}
\setlength{\floatsep}{0.1cm}

\noindent\textbf{Practical Considerations.}
The time between $t_m$ and $t_{leave}$ is called the \emph{checking period}, which spans both the marking and unlearning periods. The longer the checking period, the more certain the leaving participant is about the verification result. Nevertheless, the longer checking period also means that the leaving participant can download the global model more times than he/she should, which might be unfair to other participants.
As such, $t_{leave}$ is an important hyper-parameter that should be agreed upon among the federation. 
The $T_{enabled}$ hyper-parameter, which ends the free stage and enables unlearning, can be determined by the global training loss or accuracy. In FL, the server does not have data to compute the global loss/accuracy. Nevertheless, the server can estimate the convergence by the stability of the aggregated gradients.
It is also worth mentioning that dividing the FL process into two stages is of practical importance: it can avoid the collapse of the global model caused by the unlearning.

\noindent\textbf{System Assumption.}\label{Threat model} Following existing unlearning works \cite{liu2020federated,sommer2020towards}, we assume a trusted server with an unlearning method in place. We also assume that the local data of the involved participants remains the same in each contributed round of FL.
The server adopts partial device participation strategy in each round to motivate generating an excellent model with respect to each participant, such as choosing 10 participants among the 100 alternatives to contribute their local models each time.
Beyond the above assumption, we additionally explore the adversarial scenarios that the server and other participants are \emph{unlearning-malicious} in Section \ref{Adversarial Setting}.
\subsection{Unlearning}\label{unlearning}
Unlearning is performed by the server immediately after the completion of marking by the leaving participant.
All unlearning methods are marked by the subscription symbol $^u$ before their names. 
For a comprehensive analysis, we adopt, adapt or propose a set of comprehensive methods in this work. This gives us 7 unlearning methods in total, including 3 existing, 3 adapted and 1 newly proposed ($^u$S2U), as summarized in Table \ref{Unlearning methods summary}. 
$^u$RT and $^u$RTB are both retraining-based unlearning methods but with different retraining starting points \cite{he2021deepobliviate, ginart2019making}. $^u${CGS}, $^u${GGS} and $^u${IGS} all exploit gradient subtraction to erase the leaving data but with different gradient reconstruction strategies \cite{liu2021federaser, liu2020learn}. $^u$DP is an existing differential privacy \cite{abadi2016deep} based unlearning method \cite{shintre2019making}. Considering the high cost and negative impact of existing unlearning methods on the original task, we further propose $^u$S2U, a more efficient and friendly unlearning method that is more compatible with FL.

\subsubsection{Proposed \textbf{Scale-to-Unlearn ($^u$S2U)}}
$^u$S2U is inspired by the observation that scaling up/down the uploaded updates can substantially influence the global model \cite{bagdasaryan2020backdoor, liu2021federaser,liu2020learn}. Intuitively, scaling up/down one's local update would increase/reduce its contribution to the global model.
When unlearning is activated, $^u$S2U erases the contribution of the leaving data from the global model as follows:

\commentout{
\begin{shrinkeq}{-1ex}
\begin{equation}
\varphi \left( \vw_{t_u+1}^{(\mathsf{a})} - \vw_{t_u} \right) = \alpha\left( \vw_{t_u+1}^{(\mathsf{a})} - \vw_{t_u} \right),
\end{equation}
\end{shrinkeq}

\begin{shrinkeq}{-1ex}
\begin{equation}
\forall_{j \in \etC\diagup \mathsf{a}}  \left(\vw_{t_u+1}^{(j)} - \vw_{t_u} \right)= \beta\left(\vw_{T_{enabled}} - \vw_{t_u}\right),
\end{equation}
\end{shrinkeq}
}

\begin{equation}
  \varphi \left( \vw_{t_u+1}^{(j)} - \vw_{t_u} \right) =\begin{cases}
    \,\,\,\alpha\left( \vw_{t_u+1}^{(\mathsf{a})} - \vw_{t_u} \right), & \text{if $j$ is $\mathsf{a}$}\\
    \,\,\,\beta\left(\vw_{T_{enabled}} - \vw_{t_u}^{}\right), & \text{if $j \in \etC\diagup \mathsf{a}$}
  \end{cases}
  \label{eq:s2u}
\end{equation}

\noindent where $t_u \in [T_{enabled},T_{total}]$ is the current unlearning round (see Fig. \ref{timeline}), $\alpha \in (0, 1)$ is the down-scaling ratio, $\beta \in [1, +\infty)$ is the up-scaling ratio, and $\etC$ records the selected participants in FL at $t_u$.
Since in the unlearning-enabled stage, all local models are expected to have minimal parameter changes within a few communication rounds.
Therefore, $^u$S2U can use the global model at $T_{enabled}$ to roughly approximate other participants' local models at $t_u$: $\vw_{t_u+1}^{(j)}=\vw_{T_{enabled}}, \forall_{j \in \etC\diagup \mathsf{a}}$. Note that $^u$S2U does not need accurate approximation here.
By scaling up/down others'/$\mathsf{a}$'s local update at $t_{u}$, $^u$S2U tends to increase $\mathsf{a}$'s distance to other participants' local updates, thus actively forcing the model to unlearn $\mathsf{a}$. After unlearned by $^u$S2U, the global model is closer to other participants' local models and farther away from $\mathsf{a}$'s local model. The theoretical explanation can be found in Appendix \ref{Theoretical Explanation}.
$^u$S2U is compatible with most of the commonly used aggregation rules such as FedAvg \cite{mcmahan2017communication} and Krum \cite{blanchard2017machine}.

\begin{table*}[t]

    \centering
    \caption{A summary of unlearning methods.}
    \label{Unlearning methods summary}
    \scalebox{0.65}{
    \begin{tabular}{ccccc}
    \toprule
     &\textbf{Mechanism} &\textbf{Method}  & \textbf{Source} & \textbf{Description}         \\ 
    \midrule
    \multirow{5}{*}{Multi-Step} &  \multirow{2}{*}{Retraining} & Retraining ($^u$RT)           &    Existing          & Retrain from scratch   \\ 
    & &Retraining Breakpoint ($^u$RTB)  \cite{he2021deepobliviate, ginart2019making}        & Adapted        & Retrain from the stored intermediate model at the breakpoint                \\ \cline{2-5}
    &  \multirow{3}{*}{Gradient Subtraction} &Calibrated Gradient Subtraction($^u${CGS}) \cite{liu2021federaser} & Existing        & Subtract the calibrated unlearned gradients by leveraging others' historical updates    \\
    &  &Generated Gradient Subtraction($^u${GGS}) \cite{liu2020learn}  & Existing       & Subtract the unlearned gradients produced by a trainable dummy  generator       \\
    &  &Individual Gradient Subtraction($^u${IGS}) \cite{bourtoule2021machine, cao2015towards} & Adapted         & Subtract the leaver's gradient         \\  \midrule 
    \multirow{2}{*}{One-Step} & Covered by noise & Differential Privacy ($^u$DP)  \cite{shintre2019making}          & Adapted         & Cover the memory by introducing noise      \\ \cline{2-5}
    & \textbf{Scaling }   &\textbf{Scale-to-Unlearn  ($^u$S2U)}             & \textbf{Proposed}       & \textbf{Scale up others' gradient and scale down the leaver's gradient}     \\
    \bottomrule
    \end{tabular}}
\end{table*}

\subsubsection{Existing or Adapted Unlearning Methods}
\textbf{Retraining methods}, including Retraining ($^u$RT) and Retraining breakpoint ($^u$RTB),  retrain the global model without the leaving data.
$^u$RT reverts the global model to the starting point $\vw_0$, then retrains the model from scratch without the leaving participant $\mathsf{a}$'s local gradients. $^u$RTB is adapted from $^u$RT and it retrains the global model from a certain breakpoint $\vw_b$. 
$^u$RTB additionally requires storing the intermediate global model obtained at each communication round.

\textbf{Gradient subtraction methods}, including Calibrated Gradient Subtraction ($^u${CGS}), Generated Gradient Subtraction ($^u${GGS}) and Individual Gradient Subtraction ($^u${IGS}), erase the leaving data by subtracting the corresponding gradients.
$^u${CGS} \cite{liu2021federaser} leverages a calibration algorithm to approximate the gradients to be unlearned from other participants' historical updates. $^u${GGS} \cite{liu2020learn} deploys a trainable dummy gradient generator to produce the gradients to be unlearned. $^u${IGS} is adapted from the above two methods and it directly subtracts the local updates of the leaving participant during the next few rounds of aggregation.
Formally, these methods perform gradient subtraction as follows:
\begin{shrinkeq}{-1ex}
\begin{equation}
\varphi \left( \vw_{t+1}^{(\mathsf{a})} - \vw_{t} \right)=  - \lambda \sum_{i \in \Omega}  \left( \hat{\vw}_{i+1}^{(\mathsf{a})} -  \vw_{i} \right) , \;
\end{equation}
\end{shrinkeq}
where, $\hat{\vw}_{i+1}^{(\mathsf{a})}$ is $\mathsf{a}$'s local gradient (raw, generated or estimated) to be unlearned, $\lambda$ is a hyper-parameter balancing the unlearning of $\mathsf{a}$'s local updates and the original task, and
$\Omega$ records the rounds when $\mathsf{a}$'s gradient has been uploaded to the server.

\textbf{Differential Privacy (DP) method}, $^u$DP \cite{shintre2019making} adds noise to $\mathsf{a}$'s local updates at $t_{u} $ to smooth out the sensitive information and cover the memorization of $\mathsf{a}$'s private data in the global model:
\begin{shrinkeq}{-1ex}
\begin{equation}
\varphi  \left( \vw_{t+1}^{(\mathsf{a})} - \vw_t \right)= e^\epsilon  \left( \vw_{t+1}^{(\mathsf{a})} - \vw_t \right) + \delta, \; t = t_{u},
\end{equation}
\end{shrinkeq}
$\epsilon$ is the privacy budget, $\delta$ is a relaxation term, the smaller $\epsilon$, the more noise is added into the local model.
The central server introduces and adjusts the $\left(\epsilon,\delta\right)$ parameter pair to blur the memorization without degrading too much of the global model's performance.

\noindent\textbf{Discussion.}
Among the above 7 unlearning methods, $^u$RT and $^u$RTB are arguably the most effective yet costly unlearning methods. The 5 multi-step methods (see Table \ref{Unlearning methods summary}),  including the 2 retraining and 3 gradient subtraction methods, all need to perform unlearning for multiple communication rounds (ideally, the same number of rounds as the leaving participant's contribution in the past). As such, these methods need to store the raw, generated or estimated local/global gradients for each round. Such storage may raise new privacy concerns.
Both $^u$DP and our proposed $^u$S2U are one-step methods that only exploit the current round of gradient information. So both methods are lightweight and do not need to store the local or global gradients.
By involving noise into the gradients, $^u$DP may hurt the original task as FL heavily relies on high-quality gradients to converge.
Compared with $^u$DP, our $^u$S2U is more FL-friendly as it has minimum (or even positive) impact on other participants' local updates after aggregation.

\subsection{Verification}\label{verification}
Verification is performed by the leaving participant, consisting of two chained steps: \emph{marking} and \emph{checking}. In other words, once a marking method is determined, so does its checking method or metrics.
In Table \ref{verification summary}, we adopt, adapt or propose 5 marking methods for unlearning verification. $^v$FM and $^v$EM are our proposed non-invasive verification methods. $^v$BN inherits the backdoor-based verification in \cite{sommer2020towards}, thus also raising new security risks. $^v$ME and $^V$BF are both adapted from the deep learning intellectual property protection field \cite{uchida2017embedding, cao2019ipguard}.

\begin{table*}[!htb]
  \centering
  \caption{A summary of marking methods.}
  \label{verification summary}
  \begin{threeparttable}
  \scalebox{0.55}{
  \begin{tabular}{cccccc|ccc}
    \toprule
  \textbf{Category}                   & \textbf{Method} & \textbf{Source} & \textbf{Type}  & \textbf{Marker} & \textbf{Checking} & \textbf{\begin{tabular}[c]{@{}c@{}}Training \\ Controllability\end{tabular}} & \textbf{\begin{tabular}[c]{@{}c@{}}External \\ Data\end{tabular}} & \textbf{\begin{tabular}[c]{@{}c@{}}White-box\\ Access\end{tabular}}   \\ \midrule

  \multirow{2}{*}{Watermark} & Model Embedding($^v$ME) \cite{uchida2017embedding}    &  Adapted  & Invasive    &  Embedded bits in model parameters &  Matching rate of the extracted bits  & \faCircle    & \faCircleO & \faCircle     \\

  & BadNets($^v$BN)  \cite{sommer2020towards}            &   Existing  & Invasive  & Pixel-level backdoor trigger   & Accuracy on the backdoor samples & \faAdjust    & \faCircle & \faCircleO \\ \midrule
  \multirow{1}{*}{Fingerprint}      & Boundary Fingerprint($^v$BF) \cite{cao2019ipguard}     &   Adapted &  Invasive &    Boundary samples & Accuracy on the boundary samples & \faCircleO  & \faCircle  & \faCircleO  \\ \midrule
  \multirow{2}{*}{\textbf{Unique Memory}}  & \textbf{Forgettable Memory($^v$FM)}   &  \textbf{Proposed}  & \textbf{Non-invasive}  &  \textbf{Forgettable samples}   &  \textbf{Variance of loss on the forgettable samples}  & \faCircleO & \faCircleO  & \faCircleO    \\
& \textbf{Erroneous Memory($^v$EM)}       &    \textbf{Proposed} & \textbf{Non-invasive} &   \textbf{Erroneous samples}  & \textbf{Loss on the erroneous samples}   & \faCircleO & \faCircleO  & \faCircleO \\ 
\bottomrule                                             
  \end{tabular}}
  \begin{tablenotes}   
    \footnotesize  
    \item[]  \faCircle  ---  Required    \faAdjust  ---  Partially required    \faCircleO  ---  Not required 
  \end{tablenotes}       
    \end{threeparttable}
  \end{table*}

\subsubsection{Marking}
\label{sec:marking}
We call the marked information as `markers', a concept that is analogous to the biomarkers used in biomedical studies \cite{davis2017biology}. Intuitively, markers can be any information related to the leaving data, e.g., a subset of local samples, gradients or models.  
Table \ref{verification summary} summarizes the characteristics of the marking methods.

\noindent\textbf{Proposed Unique Memory Markers.}\label{Proposed Unique Memory Markers.}
\leavevmode
We propose to leverage the unique memories of the global model about the leaving data as effective markers.
Specifically, we propose to explore two types of unique memories: \emph{forgettable memory} and \emph{erroneous memory}\footnote{The unlearning verification effect  difference between these unique memory samples and the leaving data can be found in Appendix \ref{Verification effect difference between unique memory samples and leaving data}.}.

\begin{figure}[!htb] 
    \centering 
    \subfigure[\lab{6}]{
        \label{mnist_forgotten_data_a}
        \includegraphics[width=0.16\linewidth]{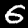}}
    \subfigure[\lab{5}]{
    \label{mnist_forgotten_data_b}
        \includegraphics[width=0.16\linewidth]{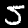}}
    \subfigure[\lab{2}]{
    \label{mnist_forgotten_data_c}
        \includegraphics[width=0.16\linewidth]{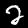}}
    \subfigure[\lab{9}]{
    \label{mnist_forgotten_data_d}
        \includegraphics[width=0.16\linewidth]{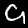}}
    \subfigure[\lab{5}]{
    \label{mnist_forgotten_data_e}
        \includegraphics[width=0.16\linewidth]{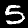}}
    \caption{Forgettable examples (markers) found by $^v$FM for one leaving participant during 10-participant FL on MNIST \cite{lecun1998gradient}.}
    \label{mnist_forgotten_data}
\end{figure}

\emph{Forgettable Memory ($^v$FM)} refers to the subset of forgettable examples by the global model. Intuitively, forgettable examples are the hardest and unique examples owned by the leaving participant, whereas unforgettable examples are easy examples shared across different participants \cite{toneva2018empirical}.
$^v$FM determines forgettable examples by the variance of their local training loss and chooses a subset of samples with the highest loss variance across several communication rounds as the markers.
Fig. \ref{mnist_forgotten_data} illustrates a few forgettable examples (i.e., markers) identified by $^v$FM from the MNIST \cite{lecun1998gradient} dataset. 
We denote the marker set found by $^v$FM for a leaving participant $\mathsf{a}$ as $D^{m}_{\mathsf{a}}$ and $D^{m}_{\mathsf{a}} \subset D_{\mathsf{a}}$.
At the marking step, $\mathsf{a}$ locally fine-tunes the model for a sufficient number of iterations to reduce the local loss variance on $D^{m}_{\mathsf{a}}$, then uploads the fine-tuned parameters to the server. Now the global model will also have relatively low loss variance on $D^{m}_{\mathsf{a}}$. During checking, $\mathsf{a}$ can monitor the global model's loss variance on $D^{m}_{\mathsf{a}}$ to verify the unlearning effect. Effective unlearning should quickly recover the high loss variation on $D^{m}_{\mathsf{a}}$.

\begin{figure}[!htb] 
    \centering 
    \subfigure[\lab{2}]{
        \includegraphics[width=0.16\linewidth]{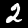}}\,\,\,
    \subfigure[\lab{7}$\rightarrow$\lab{2}]{
        \includegraphics[width=0.16\linewidth]{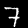}
        \includegraphics[width=0.16\linewidth]{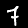}
        \includegraphics[width=0.16\linewidth]{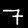}
        \includegraphics[width=0.16\linewidth]{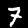}
        }
    \caption{Erroneous samples (markers) found by $^v$EM for one leaving participant during 10-participant FL on MNIST. (a): a normal image from class \lab{2}; (b): the erroneous images with majority true class \lab{7} but mostly are predicted as \lab{2}.}
    \label{mnist_semantic_data}
\end{figure}

\emph{Erroneous Memory ($^v$EM)} refers to the subset of erroneous (incorrectly predicted) samples to the global model. Intuitively, erroneous samples are likely to be the hard and rare samples uniquely owned by the leaving participant, as otherwise they should be well learned by the global model if other participants also have these samples.
As described in Algorithm \ref{alg:sdv}, $^v$EM first investigates the top $\kappa$ (\%) of the high loss samples (Line 1) and selects the majority class of erroneous samples into the marker set $D^{m}_{(\mathsf{a})}$ (Line 2). Note that the marker set has only one class (i.e., the majority class). Fig. \ref{mnist_semantic_data} shows a few erroneous MNIST samples identified by $^v$EM, in which images of class \lab{7} are misclassified as \lab{2}.
$^v$EM then relabels $D^{m}_{(\mathsf{a})}$ to its mostly predicted label by the local model $f^{(\mathsf{a})}$ (Lines 4-6) and fine-tunes the local model on the relabelled dataset to obtain a marked model $\widetilde{f}^{(\mathsf{a})}$ (Line 7). The marked model will then be uploaded to the central server to be aggregated into the global model. Fine-tuning with erroneous labels is to make the loss on the markers smaller and check if the global model can increase the loss on the markers through unlearning. Since $\mathsf{a}$ fine-tunes the local model to maintain a low loss on the $^v$EM markers during the marking process, effective unlearning should quickly recover the high losses on the markers.

\begin{algorithm}[!htb]
   \SetAlgoNoLine
   \DontPrintSemicolon
   \caption{Erroneous Memory Marking}
   \label{alg:sdv}
   \KwIn{
   The local model $f^{(\mathsf{a})}$ and private data $D_{\mathsf{a}}$ of participant $\mathsf{a}$, erroneous sample proportion $\kappa$, fine-tuning iterations $T^{(\mathsf{a})}_{ft}$.
   }
   \KwOut{
    Marked local model $\widetilde{f}^{(\mathsf{a})}$, marker dataset $D^{m}_{\mathsf{a}}$
   }
   $D^{\kappa}_{l} \leftarrow $ top $\kappa\%$ of high loss samples (and labels) in $D_\mathsf{a}$\;
  $D^{m}_{\mathsf{a}} \leftarrow $ the majority class of samples in $D^{\kappa}_{l}$\;
   $\overline{D}_{\mathsf{a}} \leftarrow D_{\mathsf{a}} \setminus D^{m}_{\mathsf{a}}$\;
   \ForEach{$(\vx,y) \in D^{m}_{\mathsf{a}}$}
   {$y$ $\leftarrow$ the most predicted label on $D^{m}_{\mathsf{a}}$ }
   $\widetilde{f}^{(\mathsf{a})}$ $\leftarrow$ fine-tune $f^{(\mathsf{a})}$ on $\overline{D}_{\mathsf{a}} \cup D^{m}_{\mathsf{a}} $ for $T^{(\mathsf{a})}_{ft}$ iterations \;
   \Return {$\widetilde{f}^{(\mathsf{a})}$, $D^{m}_{\mathsf{a}}$}
\end{algorithm}

\noindent\textbf{Existing or Adapted Marking Methods.}
\label{sec:fingerprint}
Existing watermarking methods such as parameter-based \cite{uchida2017embedding} and backdoor-based watermarking \cite{zhang2018protecting}  or fingerprinting methods \cite{cao2019ipguard} from the field of deep learning intellectual property protection can be adapted as marking methods.

For watermarking, we adopt the backdoor-based ($^v$BN) marking method from \cite{sommer2020towards} that was initially proposed for traditional machine unlearning verification. $^v$BN leverages the BadNets \cite{gu2017badnets} backdoor attack to inject trigger patterns associated with a backdoor class into the global model to verify the unlearning effect. At the marking step, $^v$BN fine-tunes the local model on backdoored data and uploads the backdoored local parameters to the server for aggregation. After fine-tuning, backdoored samples exhibit a high attack success rate on the backdoored local and global models. Effective unlearning should break the correlation between the trigger pattern and the backdoor class, i.e., lowering the attack success rate.

For fingerprinting, we adapt the Boundary Fingerprint ($^v$BF) \cite{cao2019ipguard} to find decision boundary fingerprints (markers) to verify unlearning.
$^v$BF generates adversarial examples that are close to the decision boundary to characterize 
the robustness property of the local model $f^{(\mathsf{a})}$. Arguably, the adversarial examples with relatively high and close top-2 class probabilities are boundary examples \cite{cao2019ipguard}. Therefore, before the unlearning round $t_u$ (see Fig. \ref{timeline}), $^v$BF marks the following adversarial examples as markers:
\begin{equation}
\small
 D^m_{\mathsf{a}} = \{(\vx + \sigma, y) \, \mid |f^{(\mathsf{a})}_{top-1}(\vx + \sigma) - f^{(\mathsf{a})}_{top-2}(\vx + \sigma)| \leq \gamma, (\vx, y) \in D_{\mathsf{a}}\},
\end{equation}
where, $f^{(\mathsf{a})}_{top-1}(\vx + \sigma)$ and $f^{(\mathsf{a})}_{top-2}(\vx + \sigma)$ denote the top-1 and top-2 class probabilities respectively, $\vx + \sigma$ is the PGD \cite{madry2017towards} adversarial example of $\vx$, and $\gamma \in [0, 0.1)$ is a small positive value defining how close are the two probabilities. 
At the marking step, $^v$BF first fine-tunes the local model on $D^m_{\mathsf{a}}$ to obtain a marked local model $\widetilde{f}^{(\mathsf{a})}$ which now becomes robust to $D^m_{\mathsf{a}}$ and has more smoothed boundary around the markers. The marked local model will then be uploaded to the central server and aggregated into the global model.
Effective unlearning should quickly forget the smoothed (robust) boundary around the markers (thus resulting in wrong predictions), which can be easily checked by the performance on the adversarial markers.

\noindent\textbf{Remark.}
Note that some verification methods included in this work may raise security concerns (see Appendix \ref{Security risk of $^v$BN}), or become less effective if a secure FL algorithm is implemented. One example is the backdoor-based verification proposed in \cite{sommer2020towards}. This aspect has also been considered when categorizing the verification techniques or making our recommendations.
Therefore, we categorize the marking methods in Table \ref{verification summary} into two major types: 1) \emph{invasive} methods that need to tamper with the FL process, such as modifying the global model training or injecting external data; and 2) \emph{non-invasive} methods that only need to keep track of a subset of existing data. 
The last three columns highlight the three undesired properties: training controllability, external data and white-box access. These undesired properties may introduce new security or privacy risks into FL. Invasive methods often rely on one or more of the undesired properties.
By contrast, our proposed unique memory-based methods do not require any of training controllability, external data or white-box access.

\subsubsection{Checking}\label{checking}
Checking is also performed by the leaving participant immediately after the marking. In this step, the change of the global model's performance on the markers can be used to measure the degree of unlearning. 
This process can take a few communication rounds until the leaving time $T_{leave}$. Note that the performance is directly measured on all leaving samples if markers are not used, as it did in most prior works \cite{cao2015towards,bourtoule2021machine,liu2020federated}.

Here, we consider four metrics (including accuracy, loss, influence function \cite{koh2017understanding}, and KL divergence \cite{goldberger2003efficient}) to measure the model's performance or performance change.
The accuracy and loss can be easily calculated on either the marker set or the entire leaving data.
Influence function \cite{koh2017understanding} formalizes the impact of a training sample on model prediction. We compute the influence function (IF) of all leaving samples (not just the markers) on the global model to quantify unlearning. We refer readers to \cite{koh2017understanding} for more calculation details of the influence function.
KL divergence (KL) \cite{goldberger2003efficient} measures the distributional difference between the global model's output probability distribution and an ideal with-unlearning probability distribution $\vec{\rho}$.
Arguably, the uniform distribution indicates an ideal case of unlearning, i.e., $\vec{\rho} = (\frac{1}{C}, \cdots, \frac{1}{C})$ with $C$ is the total number of classes. This gives us the following KL divergence metric on the markers for unlearning verification:
  \begin{equation}\label{eq:KL divergence}
  \small
  KL(D^{m}_{\mathsf{a}}) = \sE_{\vx \in D^{m}_{\mathsf{a}}} \Big[ f_{t}(\vx)\log\frac{f_{t}(\vx)}{\vec{\rho}}\Big], \; t \in [t_m, T_{total}).
  \end{equation}
If unlearning is effective, then the global model will not produce any meaningful predictions on the markers, resulting in a low or even zero KL divergence.

\section{Experiments}\label{evaluation}
We conduct extensive experiments with the \verifi framework to answer the key research questions (RQs) on verifiable federated unlearning defined in Section \ref{sec:intro}. All experiments are conducted on a Linux server with 4 Nvidia RTX 3090 GPUs, each with 24 GB dedicated memory, Intel Xeon processor with 16 cores and 384 GB RAM. Our code is implemented using PyTorch 1.7.1 with CUDA 11.1 and Python 3.7.

\noindent\textbf{Experimental Setup.} 
We run experiments on 7 datasets, including two popular low-resolution image classification datasets ({MNIST} \cite{lecun1998gradient} and {CIFAR-10} \cite{krizhevsky2009learning}), a speech recognition dataset (SpeechCommand \cite{warden2018speech}), two high-resolution image datasets for face ({VGGFace\_mini} \cite{parkhi2015deep}) and natural object ({ImageNet\_mini}\cite{deng2009imagenet}) recognition, and two medical image datasets for skin cancer ({ISIC} \cite{tschandl2018ham10000, codella2018skin, combalia2019bcn20000}) and COVID-19 ({COVID} \cite{dong2020interactive}) diagnoses. The datasets and corresponding models are summarized in Table \ref{model architecture}. The training data of each dataset are equally distributed to each participant, and there is no overlap between individual data. The default parameter settings (e.g., learning rate and optimizer)  are summarized in Table \ref{Parameters of FL} and Table \ref{Parameters of verification} in Appendix \ref{parameter}. The experimental setup of \verifi is summarized in Table \ref{Parameters of FL}.
The two grey highlighted hyper-parameters are for an early-stage testing (i.e., leaving immediately after unlearning is enabled) experiment only.

\begin{table}[t]
\centering
\caption{Datasets, models and test accuracies (Acc).}
\label{model architecture}
\scalebox{0.75}{
\begin{tabular}{c@{\hskip 0.05in}c@{\hskip 0.05in}c@{\hskip 0.05in}c@{\hskip 0.05in}c@{\hskip 0.05in}c}
\toprule
\textbf{Dataset} & \textbf{\#classes} & \textbf{\#samples} & \textbf{Resolution} & \textbf{Model}      & \textbf{Acc (\%)}      \\ 
\midrule
MNIST \cite{lecun1998gradient}            & 10     & 70000& 32*32        & LeNet-5 & 99.11 \\
CIFAR-10 \cite{krizhevsky2009learning}          & 10    & 60000& 32*32        & ResNet-18           &       95.37            \\
SpeechCommand \cite{warden2018speech}    & 10         & 46256 & 32*32    & CNN-LSTM          &            73.09      \\ 
ISIC \cite{tschandl2018ham10000, codella2018skin, combalia2019bcn20000}           & 4      & 8000 & 224*224        & DenseNet-121      &           68.06         \\
COVID \cite{dong2020interactive}           & 3     & 16619 & 224*224         & ResNet-18            &          88.42        \\
ImageNet\_mini \cite{deng2009imagenet}   & 10      & 13500 & 224*224       & ResNet-18          &         90.60          \\
VGGFace\_mini \cite{parkhi2015deep}    & 20        &7023  & 224*224     & ResNet-18            &      95.59            \\ 
\bottomrule
\end{tabular}}
\end{table}

\begin{table}[t]

\vspace{0.1cm} 
  \centering
   \caption{ \verifi setup. $\eta$: local learning rate; $\eta_{fl}$: global learning rate; $|B|$: local batch size; $T_{enabled}$: unlearning-enabled round; $t'_u$: unlearning round (an early-stage testing); $t'_{leave}$: leaving round (an early-stage testing); $t_m$: marking round; $t_u$: unlearning round (standard testing); $t_{leave}$: leaving round (standard testing); $T_{total}$: total round; $T_{local}$: local update epochs; $n$: number of involved participants at each round; $N$: total number of participants. }
  \label{Parameters of FL}
\scalebox{0.55}{
  \begin{tabular}{cccccccccccccc}
  \toprule
                 & \textbf{$\eta$} & \textbf{$\eta_{fl}$} & \textbf{$|B|$}  & \textbf{$T_{enabled}$} & \cellcolor[HTML]{e5e5e5}\textbf{$t'_u$} & \cellcolor[HTML]{e5e5e5}\textbf{$t'_{leave}$}  & \textbf{$t_m$} & \textbf{$t_u$} & \textbf{$t_{leave}$} & \textbf{$T_{total}$}  & \textbf{$T_{local}$} &  \textbf{$n$} & \textbf{$N$}  \\ \midrule
  MNIST         & 0.01 & 10 & 1024   & 40  &\cellcolor[HTML]{e5e5e5} 40& \cellcolor[HTML]{e5e5e5}100 & 110   & 120 & 200 & 200  & 1   & 10 & 100        \\
  CIFAR-10        & 0.1    & 1    & 128 & 40  & \cellcolor[HTML]{e5e5e5}40& \cellcolor[HTML]{e5e5e5}100 & 110   & 120 & 200& 200   & 10  & 10   & 100     \\
  SpeechCommand  & 0.1  & 1    & 256  & 25& \cellcolor[HTML]{e5e5e5}25 & \cellcolor[HTML]{e5e5e5}50 & 60& 70  & 100 & 100  & 10 & 10     & 100 \\
  ISIC          & 0.1   & 1     & 8  & 40    & \cellcolor[HTML]{e5e5e5}40 & \cellcolor[HTML]{e5e5e5}100 & 106 & 112 & 130   & 130   & 10    & 10   & 100         \\
  COVID          & 0.1   & 1  & 16  & 40    & \cellcolor[HTML]{e5e5e5}40 & \cellcolor[HTML]{e5e5e5}100 & 106 & 112 & 130   & 130    & 10    & 10   & 100         \\
  ImageNet\_mini & 0.1  & 1 & 16 & 140 & \cellcolor[HTML]{e5e5e5}140 & \cellcolor[HTML]{e5e5e5}180 & 186 & 192 &210  &210       & 10     & 10 & 100   \\
  VGGFace\_mini  & 0.1  & 1    & 16   & 240 & \cellcolor[HTML]{e5e5e5}240 & \cellcolor[HTML]{e5e5e5}300 & 306 & 312 &330  &330      & 10 & 10     & 100 \\ \bottomrule
  \end{tabular}}
  \end{table}

\subsection{Is Federated Unlearning Necessary?}

We first test what would happen if there is no unlearning but only \emph{Natural Forgetting ($^u$NF)} when a participant $\mathsf{a}$ leaves the federation. $\mathsf{a}$ is randomly chosen from all the alternative participants in FL and does not influence the final result.
We evaluate the unlearning effect of $^u$NF by comparing the global model's performance on the leaving data with that obtained via Natural Training ($^u$NT) ($\mathsf{a}$ never leaves) at the end of FL.
The results are shown in Table \ref{The performance change on the leaving data caused by $^u$NF --- CIFAR10}. It is evident that the performance differences (the \textbf{diff} columns in Table \ref{The performance change on the leaving data caused by $^u$NF --- CIFAR10}) between $^u$NF and $^u$NT are almost negligible according to all four metrics. It means that the global model still memorizes the leaving data if the participant leaves at the convergence stage.
Therefore, \emph{unlearning is necessary to actively remove information about the leaving participant's private data}.

\begin{table}[!htb]
\centering
  \caption{The absolute performance change (diff$=\lvert^u$NF$-$$^u$NT$\rvert$) on the leaving data caused by $^u$NF.}
  \label{The performance change on the leaving data caused by $^u$NF --- CIFAR10}
  \scalebox{0.7}{
\begin{tabular}{c|c|ccc|ccc}
\toprule
\multirow{2}{*}{\textbf{Dataset}} & \multirow{2}{*}{\textbf{Metrics}} &  \multicolumn{3}{c|}{\textbf{At the Leaving Round}} & \multicolumn{3}{c}{\textbf{At the End of FL}} \\ \cline{3-8}
   &&   $^u$NT & $^u$NF& \textbf{diff}   & $^u$NT & $^u$NF & \textbf{diff}   \\  \midrule
\multirow{4}{*}{CIFAR10} & Acc (\%)  & 77 & 77.8&\cellcolor[HTML]{e5e5e5}0.8 & 87 & 85&\cellcolor[HTML]{e5e5e5}2  \\
 & Loss  & 0.14 & 0.14&\cellcolor[HTML]{e5e5e5}0.0 & 0.08 & 0.08 & \cellcolor[HTML]{e5e5e5}0.0 \\
 & KL  & 6.75 & 6.92&\cellcolor[HTML]{e5e5e5}0.17 & 8.31 & 8.47&\cellcolor[HTML]{e5e5e5}0.16  \\
 & IF  & 9.21e-7 & 5.67e-5&\cellcolor[HTML]{e5e5e5}5.58e-5 & 1.89e-7 & 1.98e-5 &\cellcolor[HTML]{e5e5e5}1.98e-5 \\ \midrule
\multirow{4}{*}{\begin{tabular}[c]{@{}l@{}}Speech\\ Command\end{tabular}} & Acc (\%)  & 64.61 & 64.67&\cellcolor[HTML]{e5e5e5}0.06 & 64.73 & 67.21& \cellcolor[HTML]{e5e5e5}2.48 \\
 & Loss & 0.50 & 0.52 & \cellcolor[HTML]{e5e5e5}0.02& 0.48 & 0.5& \cellcolor[HTML]{e5e5e5}0.02\\
 & KL  & 2.1 & 2.08& \cellcolor[HTML]{e5e5e5}0.02& 2.46 & 2.24&\cellcolor[HTML]{e5e5e5}0.22  \\
 & IF  & -0.007 & -0.006 & \cellcolor[HTML]{e5e5e5}0.001 & 0.006 & 0.006 & \cellcolor[HTML]{e5e5e5}0.0 \\ \midrule
\multirow{4}{*}{Covid} & Acc (\%) & 68.18 & 65.15 & \cellcolor[HTML]{e5e5e5}3.03& 81.82 & 80.3& \cellcolor[HTML]{e5e5e5}1.52 \\
 & Loss  & 0.49 & 0.36& \cellcolor[HTML]{e5e5e5}0.13 & 0.26 & 0.3& \cellcolor[HTML]{e5e5e5}0.04 \\
 & KL & 0.31 & 0.66 & \cellcolor[HTML]{e5e5e5}0.35& 1.12 & 0.99 &\cellcolor[HTML]{e5e5e5}0.13 \\
 & IF  & 1.34e-5 & 0.03& \cellcolor[HTML]{e5e5e5}0.03& 0.001 & 0.001&\cellcolor[HTML]{e5e5e5}0.0 \\ \midrule
\multirow{4}{*}{\begin{tabular}[c]{@{}l@{}}VGGFace\\   \_mini\end{tabular}} & Acc (\%)  & 57.14 & 69.64& \cellcolor[HTML]{e5e5e5}12.5& 78.57 & 76.79&\cellcolor[HTML]{e5e5e5}1.78 \\
 & Loss & 0.66 & 0.54& \cellcolor[HTML]{e5e5e5}0.12& 0.63 & 0.49 &\cellcolor[HTML]{e5e5e5}0.14 \\
 & KL  & 3.03 & 3.81& \cellcolor[HTML]{e5e5e5}0.78& 3.47 & 3.99 &\cellcolor[HTML]{e5e5e5}0.52 \\
 & IF  & 0.085 & 0.247& \cellcolor[HTML]{e5e5e5}0.162& 0.029 & 0.073&\cellcolor[HTML]{e5e5e5}0.044 \\ \bottomrule
\end{tabular}
}
\end{table}

\subsection{Are Markers Necessary for Verification?}

\begin{figure*}[t]
\centering
  \includegraphics[width=1.0\linewidth]{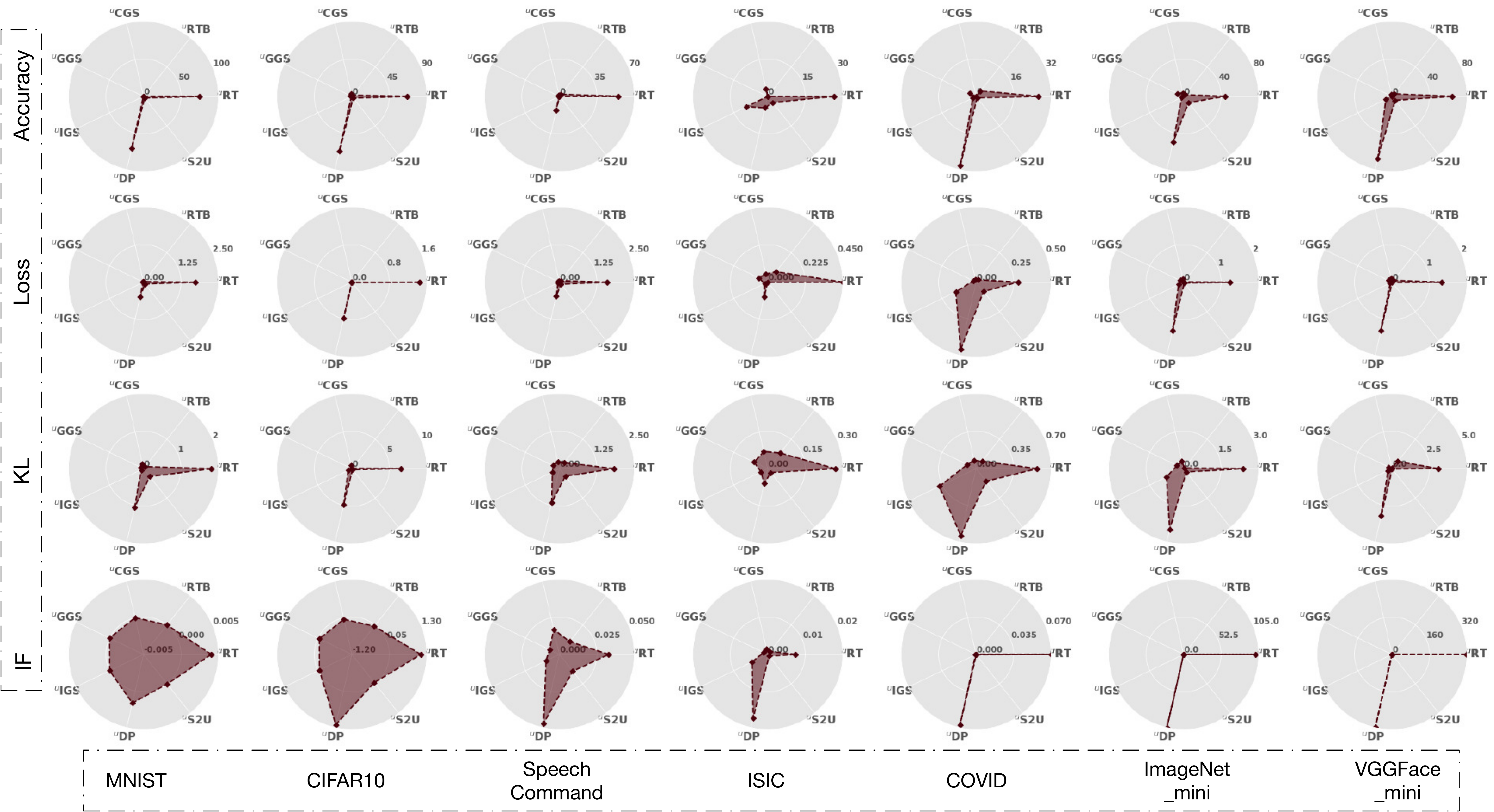}

  \caption{Verifying the unlearning effect using only the 4 metrics (rows) computed on the leaving data. Each radar chart has 7 dimensions corresponding to the 7 unlearning methods, with each dimension showing the metric difference before and after (after minus before) unlearning. Each column of radar charts correspond to one dataset. Failed verification occurs at dimensions with almost zero difference before and after unlearning.
  }
  \label{Unlearning performance comparison}
 \end{figure*}

To answer the question, we run experiments to verify the different unlearning effects of the 7 unlearning methods using only the checking metrics without any marking methods (markers). Intuitively, if the checking metrics alone can properly identify the difference before and after unlearning, then specialized markers are unnecessary. For each of the 4 metrics (i.e., accuracy, loss, KL and IF), we compute its difference before (at $t_m$) and after (at $t_u$) unlearning on the leaving data $D_{\mathsf{a}}$. Take accuracy as an example, the metric difference is computed as follows:
     \begin{equation}\label{eq:metric_diff}
         Acc_{diff}(D_{\mathsf{a}}) = \lvert Acc_{t_m}(D_{\mathsf{a}}) - Acc_{t_u}(D_{\mathsf{a}}) \rvert.
     \end{equation}
Similarly, we can define other three metrics:  $Loss_{diff}$, $KL_{diff}$, and $IF_{diff}$.

We plot the 4 metric differences for all 7 unlearning methods on each dataset in Fig. \ref{Unlearning performance comparison}. Large metric differences (large covered area in a radar chart) indicate successful verification. For a given metric, if it successfully verifies the difference before and after unlearning across different datasets, it can be regarded as an effective metric for federated unlearning verification. Unfortunately, as shown in Fig. \ref{Unlearning performance comparison}, we find that, in general, none of the metrics can effectively verify the unlearning effects of all unlearning methods. Furthermore, among the 7 unlearning methods, \emph{$^u$DP and $^u$RT are relatively easier to verify by any of the 4 metrics}. This means that we don't need sophisticated verification methods if $^u$DP or $^u$RT is adopted as the unlearning method. While this result is encouraging, the two unlearning methods also have their own weaknesses. For instance, \emph{$^u$RT is very costly and $^u$DP causes the most performance drop} among the 7 unlearning methods (see Table \ref{Cost of unlearning}).
We have also tested two naive methods for verification: model parameter difference and privacy leakage difference in Appendix \ref{sec:parameter_distance} and \ref{sec:membership_inference}, respectively. The results show that the parameter difference (measured by Euclidean distance or Cosine similarity) of the global model before and after unlearning is also insufficient for verifying the unlearning methods, except $^u$DP, $^u$RT and our $^u$S2U.
Furthermore, from the perspective of privacy leakage \cite{chen2021machine}, i.e., the success rate of membership inference of the leaving data,
even $^u$RT cannot verify (e.g., not having a noticeable lower success rate than $^u$NT) due to the contributions of other participants.
Overall, we conclude that \emph{many (5/7) of the unlearning methods may not be properly verified by the 4 metrics without specialized markers. More effective unlearning methods like $^u$DP and $^u$RT have certain weaknesses for practical usage}.

\subsection{Federated Unlearning Verification with Markers}\label{sec:markers_experiment}

\begin{figure*}[t]
  \centering
 \includegraphics[width=1.0\linewidth]{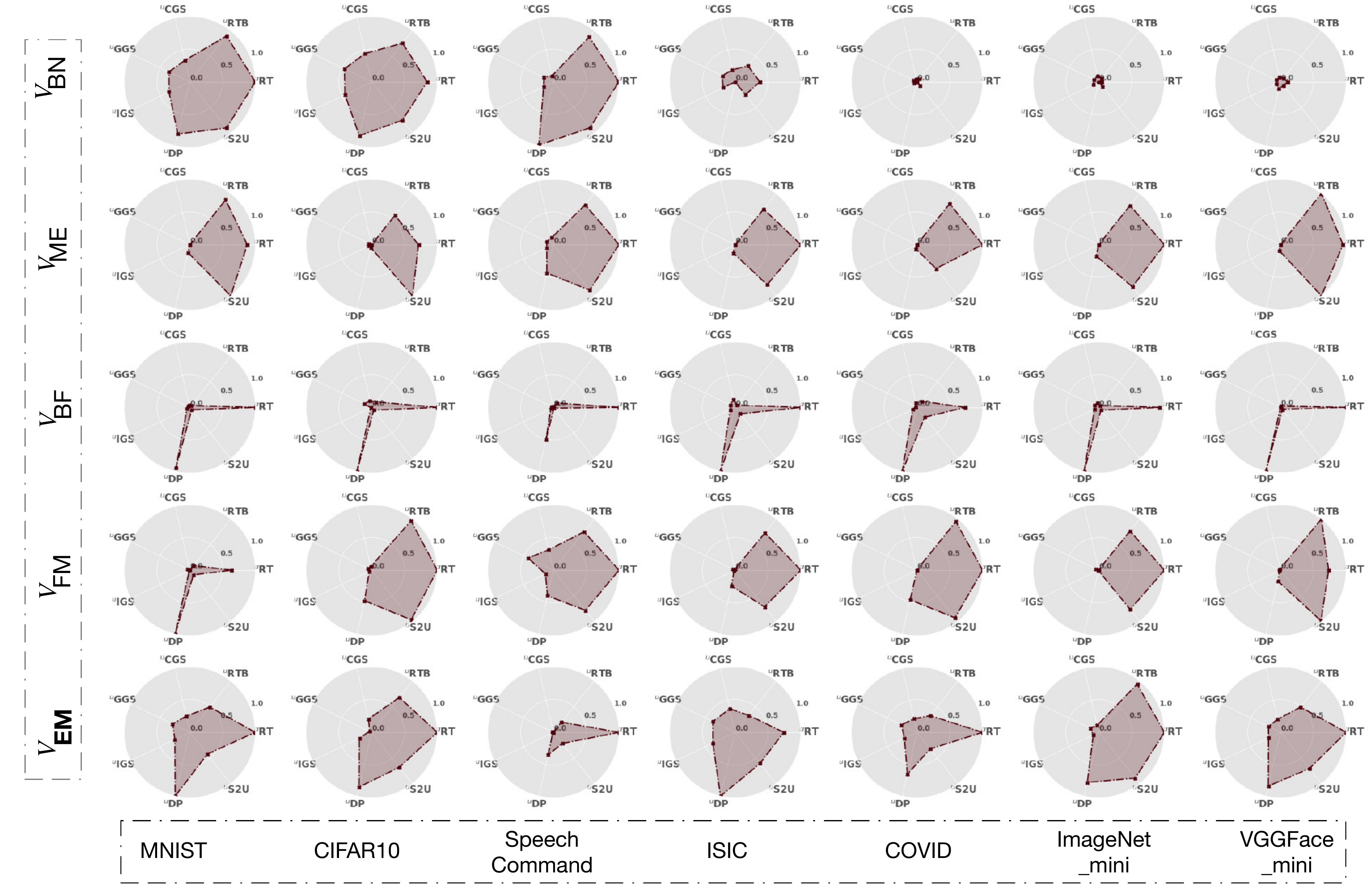}
  \caption{Verifying the unlearning effect using 5 types of markers with each row representing one type of markers and each column corresponding to one dataset. The 7 dimensions of each radar chart correspond to the 7 unlearning methods, with each dimension showing the normalized metric difference (with log transformation) before and after (after minus before) unlearning.
  The most effective metric for markers $^v$BN, $^v$ME,  and $^v$BF is accuracy, while the most effective metrics for our memory-based markers $^v$EM and $^v$FM are loss and loss variance, respectively.}
  \label{Verification performance comparison}
  \end{figure*}
  
Here, we verify the unlearning effect of the 7 unlearning methods, with the 5 marking methods (markers). Similarly, we compute the metric difference of the global model on the marker set $D^M_\mathsf{a}$ before and after unlearning following \eqref{eq:metric_diff}. Due to space limitations, we only show the most effective metric for each type of marker\footnote{The $^v$BN result is normalized with the ideally maximum accuracy gap 100\%, others are normalized according to the maximum gap value, owing to the unsuccessful backdoor-watermark injection in the big datasets.}. The results are visualized in Fig. \ref{Verification performance comparison}. A valid marking method should recognize $^u$RT as the most effective unlearning method as $^u$RT is the golden standard (i.e., the best unlearning one could achieve). And our $^u$S2U should be more effective than $^u$CGS, $^u$IGS and $^u$GGS, since it not only downscales the leaving gradients but also upscales other participants' gradients. An ideal marking method should be able to distinguish the different strengths of the unlearning methods.

\noindent\textbf{The most effective verification method.} In general, \emph{our proposed \textbf{$^v$EM} demonstrates better verification ability than $^v$FM, $^v$ME and $^v$BN \footnote{Specifically, to avoid the instant performance change on the backdoor-based watermarking method, we take the median performance during $[t_m,t_m+2]$ as the result on the markers at $t_m$.}, while $^v$BF ranks the last}. 
Particularly, $^v$EM markers could always distinguish (showing larger metric differences) stronger unlearning ($^u$RT, $^u$RTB, $^u$DP and $^u$S2U) from the mild ones ($^u$CGS, $^u$IGS and $^u$GGS) on all datasets. 
By contrast, $^v$FM could not effectively distinguish the unlearning effect of $^u$S2U and $^u$RTB on MNIST.
$^v$ME (injecting a bit string into the model parameter space) fails to verify $^u$DP as it is not sensitive to the noise of $^u$DP. 
Meanwhile, backdoor-based markers like $^v$BN fail to mark the global model on the high-resolution datasets (the accuracy on $^v$BN markers of the ISIC4, COVID, ImageNet\_mini and VGGFace\_mini datasets is similar to random guessing), and thus lose the verification ability. This result indicates that \emph{the performance of invasive marking methods cannot be guaranteed in practice}.
$^v$BF can only distinguish the unlearning effect of $^u$RT and $^u$DP as other unlearning methods will not cause significant change on the decision boundaries.

\noindent\textbf{The most effective unlearning methods.}
By examining the verified unlearning effect by the most effective marker $^v$EM, we can also cross-validate the effectiveness of the 7 unlearning methods. In general, \textbf{$^u$RT}, \textbf{$^u$RTB}, \textbf{$^u$DP} and our proposed \textbf{$^u$S2U} demonstrate more effective unlearning effects than the other 3 unlearning methods. Note that, as a completely retraining method, $^u$RT is arguably the most effective unlearning one could achieve and it is not surprising that $^u$RT demonstrates better unlearning effects than $^u$RTB, $^u$DP and $^u$S2U on nearly all datasets. The other three gradient subtraction based unlearning methods ($^u$CGS, $^u$IGS and $^u$GGS) exhibit limited unlearning effectiveness on the $^v$EM markers.

\noindent\textbf{Robustness to the byzantine-robust aggregation rules.}
We investigate the robustness of the most effective marker $^v$EM and two invasive markers $^v$BN and $^v$ME when the server adopts different aggregation rules. 
The results on CIFAR-10 dataset are reported in Table \ref{verification robustness}.
It is clear that the metric difference identified by $^v$BN and $^v$ME drops drastically when robust aggregation rules like Krum and Median are used at the server side\footnote{The two robust aggregation rules are widely applied and can be modified and combined to form other aggregation rules, such as Bulyan \cite{guerraoui2018hidden} and Trimmed Mean \cite{yin2018byzantine}.}. By contrast, our $^v$EM can maintain a stable difference, i.e., it is reasonably robust to the byzantine-robust aggregation rules.

\begin{table}[!htb]
  \centering
\caption{Verification robustness to different aggregation rules on CIFAR10 dataset with $^v$RT unlearning. `diff': absolute metric difference before and after $^v$RT.}
\label{verification robustness}
\scalebox{0.95}{
  \begin{tabular}{cccccc}
   \toprule
\textbf{Verification} & \textbf{Rule} & \textbf{Metrics} & \textbf{Before} & \textbf{After} & \textbf{diff} \\  \midrule
\multirow{3}{*}{$^v$BN} & FedAvg & Accuracy  & 84.8 & 0.0  &  \cellcolor[HTML]{67000d}\color{white}84.8   \\
 & Krum & Accuracy & 6.2  & 0.0   &  \cellcolor[HTML]{fff5f0}6.2    \\
 & Median & Accuracy  & 9.8  & 0.0  &  \cellcolor[HTML]{ffeee6}9.8   \\  \midrule
\multirow{3}{*}{$^v$ME} & FedAvg & Accuracy  & 71.88 & 48.44  & \cellcolor[HTML]{67000d}\color{white}23.44   \\
 &  Krum & Accuracy &  39.06 & 48.44  & \cellcolor[HTML]{faeae1}9.38  \\
 & Median & Accuracy  & 40.62  & 48.44 & \cellcolor[HTML]{f6f7f7}7.82    \\   \midrule
\multirow{3}{*}{$^v$EM} & FedAvg & Loss  & 12.3  & 28.1 & \cellcolor[HTML]{67000d}\color{white}15.8   \\
 &  Krum & Loss  & 14.61 & 25.69  & \cellcolor[HTML]{d92523}11.08   \\
 & Median & Loss  & 12.14  & 27.09 & \cellcolor[HTML]{800610}\color{white}14.95 \\  \bottomrule
\end{tabular}}
  \end{table}

\noindent\textbf{Unlearning cost.}
Here, we investigate the cost of different unlearning methods.
As shown in Table \ref{Cost of unlearning}, $^u$S2U demonstrates the least overall computational overhead and minor influence on the initial FL task.
Besides, $^u$RT needs the most time as it retrains from scratch. $^u$RTB needs the most space as it saves the intermediate models. $^u$CGS is both time- and space-consuming as the gradients to be unlearned need to be calibrated based on other participants' gradients. 
By contrast, the time/space cost of $^u$GGS and $^u$IGS are less than $^u$CGS as they directly construct the leaving gradients without using other participants' models. $^u$DP needs little time/storage cost by simply adding noises while causing most performance drop. 
Apart from $^u$DP and $^u$RT, other unlearning methods hardly degrade the global model's performance at the end of FL\footnote{The minor negative impact of $^u$GGS, $^u$IGS and $^u$CGS on the original task can be owned to the hyperparameter $\lambda$.}.

\noindent\textbf{Verification cost.}
Here, we study the time cost, storage cost and negative impact on the original FL task for different verification methods. The results are reported in Table \ref{Cost of verification --- CIFAR10}. All marking methods cause tolerable performance drop (either in terms of loss or accuracy). Among the 5 marking methods, $^v$FM shows less time/space cost
than $^v$EM and $^v$ME, while $^v$BN is the most time-consuming marking method as it needs to inject a backdoor watermark into the global model. $^v$BF also requires much time/space to save and generate the boundary fingerprints.
The verification method with less time overhead would produce the acceptable time delay in the large-scale practical FL system.

\begin{table}[t]
\centering
\vspace{0.1cm} 
\caption{Unlearning costs measured on CIFAR10 dataset.}
\label{Cost of unlearning}
\scalebox{0.95}{
    \begin{tabular}{ccccc}
      \toprule
    \textbf{} & \textbf{Time(s)} & \textbf{Space(MB)}& \textbf{$T_{total}\Delta$Acc}  & \textbf{$T_{total}\Delta$Loss} \\ \midrule
    $^u$RT        & \cellcolor[HTML]{67000d}\color{white}8157.91          & \cellcolor[HTML]{fb694a}44                      & \cellcolor[HTML]{f9efe9}-4.17                & \cellcolor[HTML]{fee8dd}0.14            \\
    $^u$RTB       & \cellcolor[HTML]{e93529}\color{white}929.87           & \cellcolor[HTML]{6f020e}\color{white}1760           & \cellcolor[HTML]{f6f7f7}0.18               & \cellcolor[HTML]{fff5f0}0.0                      \\
    $^u$CGS       & \cellcolor[HTML]{d11e1f}\color{white}1516.37          & \cellcolor[HTML]{67000d}\color{white}1936                 & \cellcolor[HTML]{f7f5f4}-0.82               & \cellcolor[HTML]{fff0e9}0.05               \\
    $^u$GGS       & \cellcolor[HTML]{fc8e6e}186.34           & \cellcolor[HTML]{fff5f0}0                      & \cellcolor[HTML]{f7f5f4}-0.8              & \cellcolor[HTML]{fff2ec}0.03              \\
    $^u$IGS       & \cellcolor[HTML]{fedfd0}37.9             & \cellcolor[HTML]{de2b25}\color{white}176                    & \cellcolor[HTML]{f6f7f7}0.13              & \cellcolor[HTML]{fff4ee}0.02                 \\
    $^u$DP        & \cellcolor[HTML]{fff5f0}17.1             & \cellcolor[HTML]{fff5f0}0                  & \cellcolor[HTML]{faeae1}-6.26            & \cellcolor[HTML]{fee0d2}0.22               \\
    $^u$S2U       & \cellcolor[HTML]{ffeee7}21.61            & \cellcolor[HTML]{fb694a}44                    & \cellcolor[HTML]{f6f7f7}0.05                & \cellcolor[HTML]{fff4ef}0.01        \\ \bottomrule   
    \end{tabular}}
    \end{table}

\begin{table}[t]
  \centering
\caption{Verification costs measured on CIFAR10 dataset.}
\label{Cost of verification --- CIFAR10}
\scalebox{0.95}{
  \begin{tabular}{ccccc}
  \toprule
  \textbf{} & \textbf{Time(s)} & \textbf{Space(KB)} & \textbf{$T_{total}\Delta$Acc} & \textbf{$T_{total}\Delta$Loss} \\  \midrule
  $^v$BN        & \cellcolor[HTML]{67000d}\color{white}263.5            & \cellcolor[HTML]{fff5f0}4                  &  \cellcolor[HTML]{f7f6f6}   -0.53         &   \cellcolor[HTML]{f7f5f4}     0.02       \\
  $^v$ME        & \cellcolor[HTML]{fff5f0}90.2             & \cellcolor[HTML]{fdccb8}12                 &   \cellcolor[HTML]{f8f4f2} -1.64          &     \cellcolor[HTML]{f8f1ed} 0.08         \\
  $^v$BF        & \cellcolor[HTML]{fcab8f}142.1            & \cellcolor[HTML]{67000d}\color{white}1233               &  \cellcolor[HTML]{f8f4f2}     -1.54       &    \cellcolor[HTML]{f8f2ef}  0.06         \\
  \textbf{$^v$EM}        & \cellcolor[HTML]{ffece3}99.6             & \cellcolor[HTML]{fff5f0}4                  &   \cellcolor[HTML]{f8f2ef}   -2.82        &   \cellcolor[HTML]{fee8dd}    0.15        \\
  \textbf{$^v$FM}        & \cellcolor[HTML]{fbfaf9}10.26           & \cellcolor[HTML]{fff5f0}4                  &    \cellcolor[HTML]{f8f2ef}   -2.73     &    \cellcolor[HTML]{fee8dd} 0.13   \\ \bottomrule        
  \end{tabular}}
  \end{table}

\noindent\textbf{Correlation between the markers and the leaving data.}\label{Correlation}
The unlearning effect is more pronounced on the markers than on the leaving data as the markers are specially designed to serve this purpose. This raises a natural question \emph{to what extent can the markers represent the leaving data}? 
To answer this question, we analyze the correlation between the global model's performance on the markers and on the leaving data when adopting $^u$RT during $[t_u-n,t_u+n]$ as an example. 
In this experiment, $n$ is set to 10 on MNIST, CIFAR-10 and SpeechComamnd datasets, and 6 on other datasets.
As shown in Fig. \ref{Correlation}, the performance trends on the markers (except $^v$BN markers) and the leaving data show a strong correlation before and after unlearning. This confirms that unlearning the markers can largely reflect the degree to which the server is unlearning the leaving data.

\begin{figure}[!htb]
\centering
\includegraphics[width=1\linewidth]{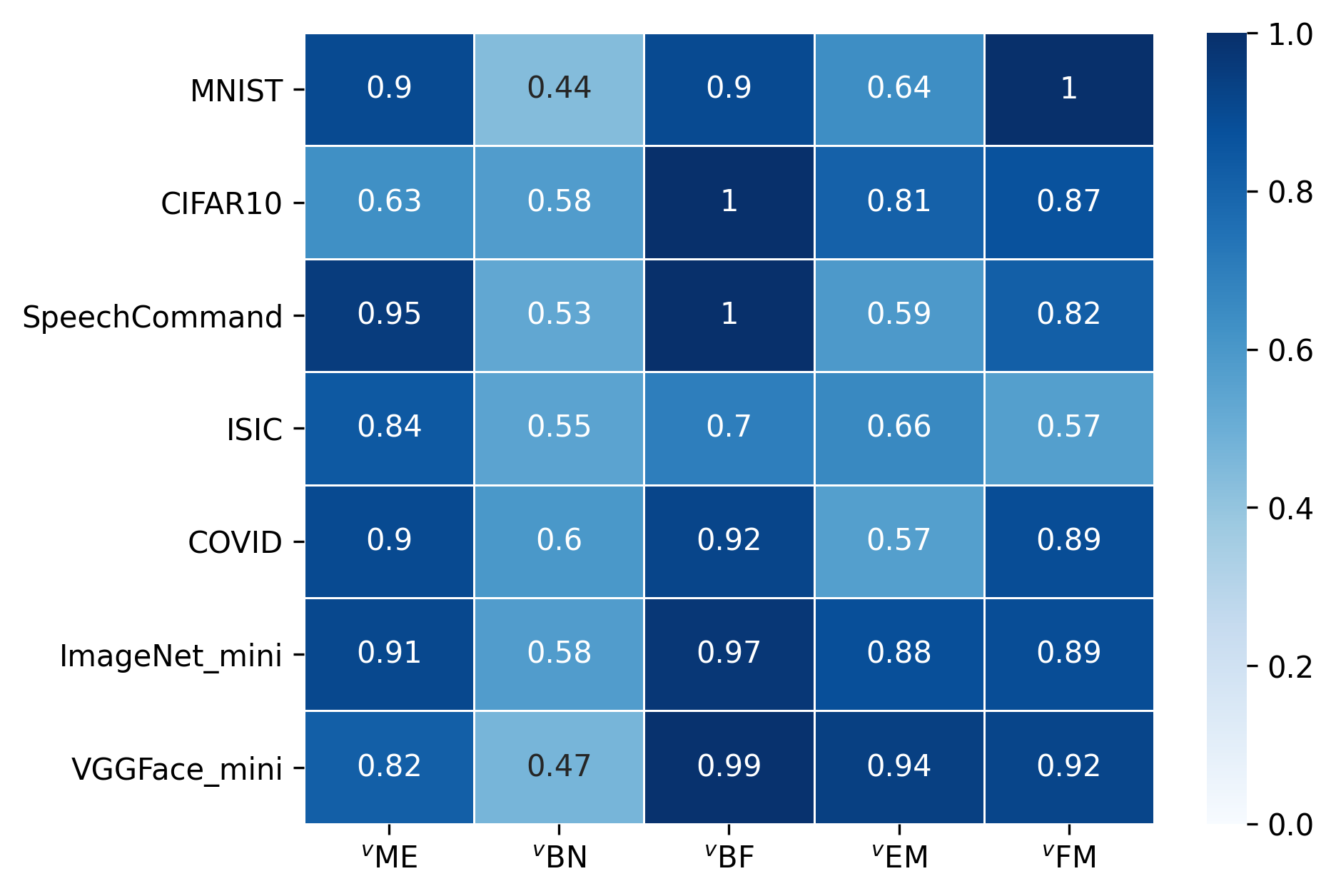}
\caption{Correlation between unlearning the markers vs. unlearning the leaving data.}
\label{Correlation}
\end{figure}

\subsection{Unlearning-Verification: The Combinations}

\begin{figure}[t]
  \centering
   \subfigure[CIFAR-10]{
        \label{uploading model with marking}
        \includegraphics[width=0.48\linewidth]{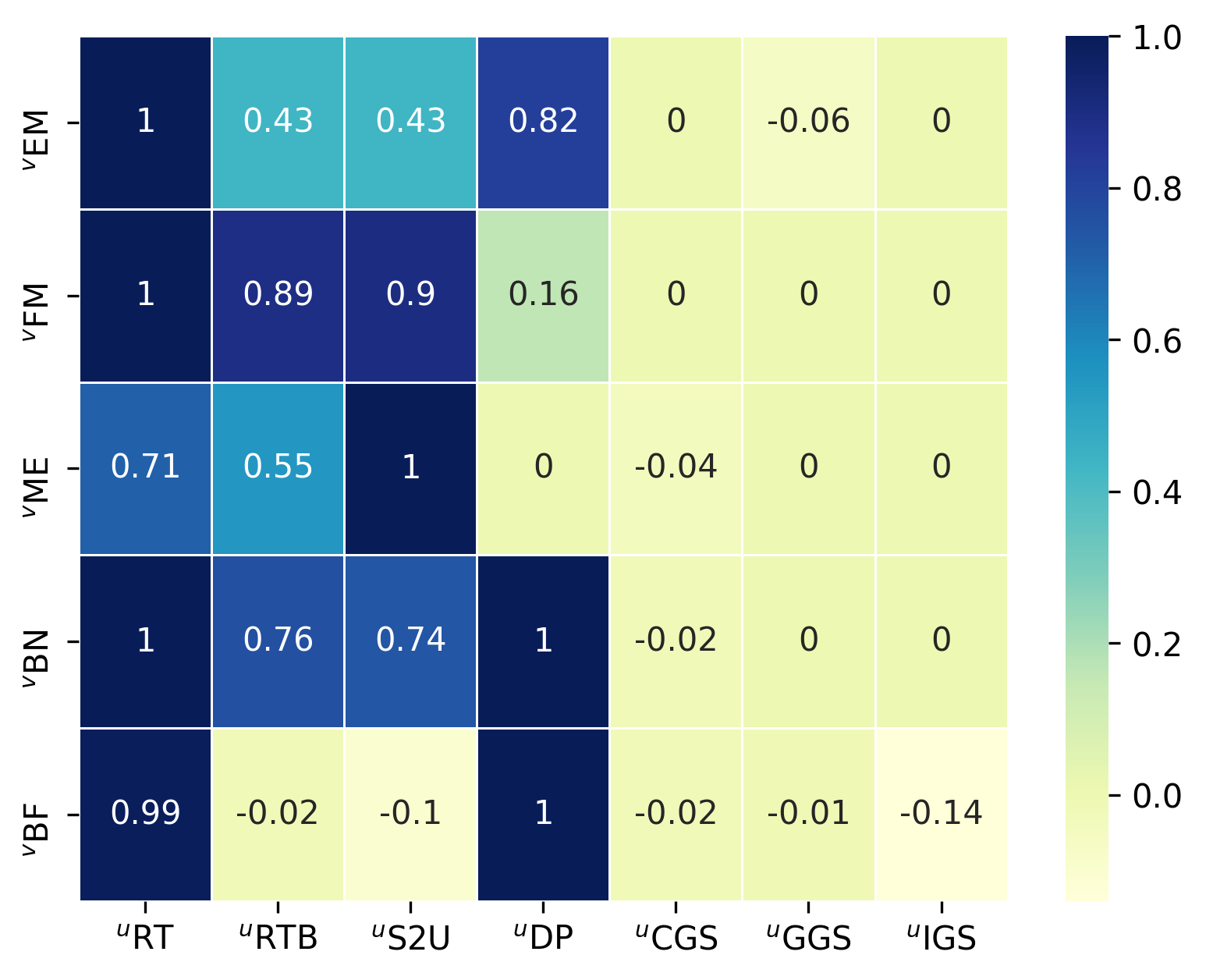}}
        \subfigure[SpeechCommand]{
        \label{uploading model without marking}
        \includegraphics[width=0.48\linewidth]{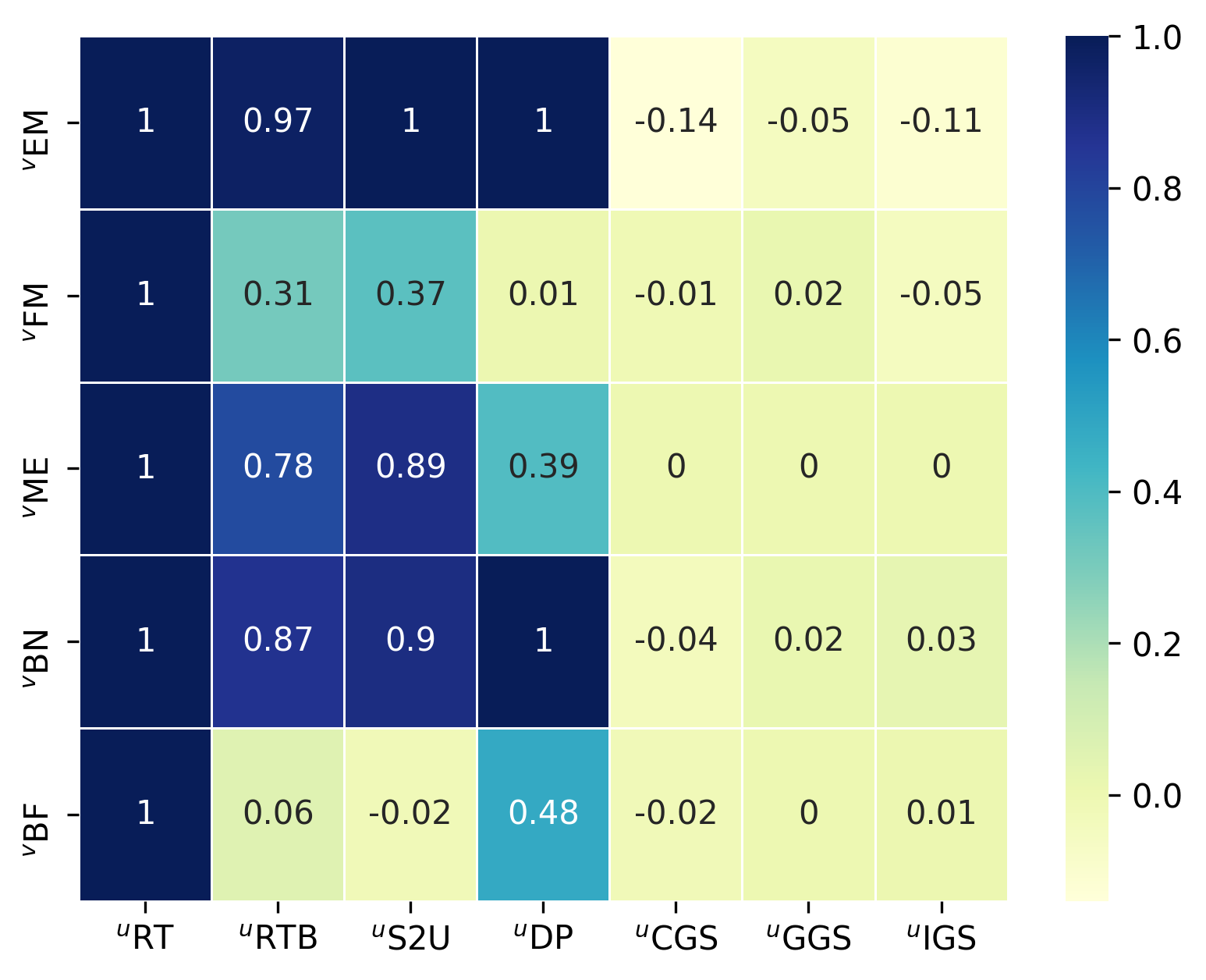}}
  \caption{The unlearning effect of 7 unlearning methods (columns) verified by 5 marking methods (rows). The unlearning effect is the normalized (max: 1, min: 0) metric difference within each marking method (row): in each row, the maximum verifiable effect is 1 while the  minimum is 0. The higher the normalized score, the better the unlearning-verification method.}
  \label{Verifiable unlearning heatmap}
  \end{figure}

The verification method goes with the unlearning method. Fig. \ref{Verifiable unlearning heatmap} shows the normalized verifiable unlearning effect of all the combinations of 7 unlearning methods and 5 verification methods on CIFAR-10 and SpeechCommand datasets. Each cell is associated with an unlearning method and a verification method. The blue cells highlight the best verifiable unlearning effect. 
Combining our analyses above, we obtain the following findings.
 A general effectiveness ranking of the unlearning methods is:  $^u$RT > $^u$RTB > $^u$S2U > $^u$DP > $^u$CGS $\approx$ $^u$GGS $\approx$ $^u$IGS. Considering the high cost of $^u$RT and $^u$RTB, and the negative impact of $^u$DP on FL, \emph{it leaves our proposed $^u$S2U to be the most promising unlearning method for its relatively higher effectiveness, higher efficiency, less negative influence on FL and higher verifiability}. It is thus promising for future work to explore similar unlearning strategies or improve $^u$DP for more effective, efficient, harmless and verifiable federated unlearning.

As for the verification methods, the deeper blue colored cells are more effective. So, the general ranking is: $^v$EM > $^v$FM > $^v$ME > $^v$BF > $^v$BN. We put $^v$ME, $^v$BF and $^v$BN to the end of the list is because they are all invasive methods that may introduce new security risks into FL (see Appendix \ref{Security risk of $^v$BN}). This makes our proposed $^v$EM the most promising verification method. \emph{The combination of our proposed $^u$S2U with $^v$EM verification is the most promising federated unlearning-verification strategy.} If $^u$DP can be improved for FL, then the $^u$DP-$^v$EM can also be an effective combination.

\section{More Explorations}\label{discussion}
  
\subsection{{Adversarial Setting}}\label{Adversarial Setting}

We also use \verifi to analyze a challenging adversarial setting where the attacker (the unlearning-malicious server or participant) may store the global model before participant $\mathsf{a}$ leaves and then restore $\mathsf{a}$'s memory after $\mathsf{a}$ leaves.
This will compromise $\mathsf{a}$'s privacy.
Since the unlearning-malicious server would not easily retreat at the cost of losing the excellent model updates from others, we then focus on the unlearning-malicious participant setting.
We take the verification method $^v$ME as an example, which checks unlearning based on the extracted bits from the model parameters. 
The successfully marked model by $^v$ME would maintain a high and stable accuracy on the $^v$ME markers.
We assume the server implements the ideal unlearning method $^u$RT which could effectively erase the memory about $\mathsf{a}$'s leaving data and markers.

\begin{figure}[!htb]
  \centering
  \subfigure[with marking]{
\label{uploading model with marking}
\includegraphics[width=0.48\linewidth]{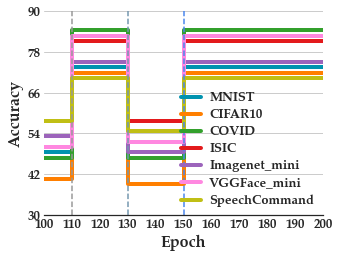}}
\subfigure[without marking]{
\label{uploading model without marking}
\includegraphics[width=0.48\linewidth]{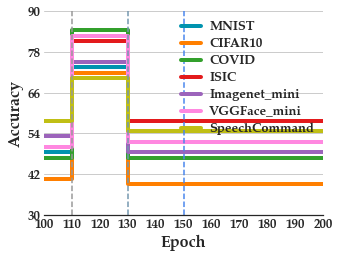}}
  \caption{The attacker uploads the historical global model (in and out of the marking stage) to attack the effectiveness of the unlearning.  The three vertical lines mark the time of marking  $t_m$, unlearning $t_u$, and attack, respectively. }
  \label{Adaptive attacker}
\end{figure}

Fig. \ref{Adaptive attacker} shows the different results when the attacker could capture and upload the global model in and out of the marking stage.
As shown in Fig. \ref{uploading model with marking}, $^u$RT decreases the accuracy on the markers at the unlearning round.
However, the accuracy on the markers arises after the attack.
In \verifi, the leaver continuously tracks the global model to check unlearning for a while, not instantly.  
Therefore, the performance rise on the markers can be checked by the leaver, and the leaver would deem the unlearning invalid.
However, if the attacker only captures the global model out of the marking stage in Fig. \ref{uploading model without marking}, 
the accuracy on the markers would not change. Thus, the leaver would deem the unlearning effective. 
Admittedly, \verifi can only detect the deceived unlearning situation at a certain probability. 
Fortunately, the retrievable model out of the marking period would maintain a longer time span and a larger difference from the attacker's local model at the previous round, which would raise more attention. Thus, we can improve the accuracy of cheated unlearning checking by analyzing the similarity between the models of the adjacent rounds. 
This is an exciting problem that is worth further exploration.

\subsection{{More Unlearning and Verification Parameters}}\label{Parameter Analysis}
We make a comprehensive analysis to explore the parameter influence in \verifi. We take the mature verification method --- $^v$BN as an example, the concrete result can be found below.

\begin{figure}[!htb] 
\setlength{\abovecaptionskip}{-0.05cm}
  \centering
  \subfigure[10-th round]{ 
      \label{epoch 10}
      \includegraphics[width=0.31\linewidth]{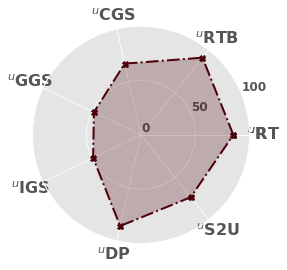}}
  \subfigure[110-th round]{
        \label{epoch 110}
        \includegraphics[width=0.31\linewidth]{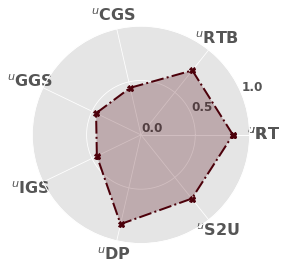}}
    \subfigure[210-th round]{
          \label{epoch 210}
          \includegraphics[width=0.31\linewidth]{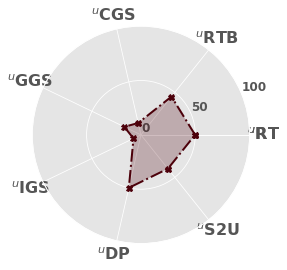}}
  \caption{Leaving time influence on CIFAR10 dataset.}
  \label{Withdrawl epoch influence in BN}
\end{figure}

\noindent\textbf{Influence of the marking time to the marking effect: }
Fig. \ref{epoch 10} $\sim$ Fig. \ref{epoch 210} presents the unlearning verification results when the marking time is respectively 10-th round (earlier), 110-th round (proper) and 210-th round (later).
With the marking time getting later, the success probability of marking decreases, and the performance change caused by unlearning reduces. 
As for the reason, when the model has converged to a stable state, injecting the watermark into the global model gets harder. Meanwhile, the unlearning effect on the markers at $t_u$ degrades, increasing the difficulty of distinguishing the authentic unlearning effectiveness.
Thus, it's better to activate unlearning and verification when the global approximately converges.

\begin{figure}[!htb] 
\setlength{\abovecaptionskip}{-0.05cm}
\vspace{-0.45cm} 
  \centering
  \subfigure[(5, 0.0)]{
      \label{size 5 trans 0.0}
      \includegraphics[width=0.31\linewidth]{./figure/radar/cifar_badnet_acc_radar.png}}
  \subfigure[(2, 0.0)]{
      \label{size 2}
      \includegraphics[width=0.31\linewidth]{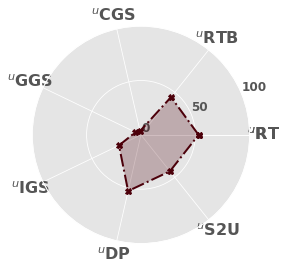}}
  \subfigure[(5, 0.6)]{ 
      \label{trans 0.6}
      \includegraphics[width=0.31\linewidth]{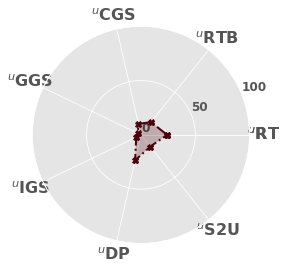}}
  \caption{Marking parameter (size, transparency) influence on CIFAR10 dataset. (a) shows a stronger backdoor with big size 5*5 and low transparency 0.0, (b) and (c) show a weaker backdoor with small size 2*2 or high transparency 0.6.}
  \label{Parameter influence in BN}
\end{figure}

\noindent\textbf{Influence of marking parameters:}
Different marking parameters inevitably cause unevenly marking effect, and further influence the unlearning verification result.
We take the size and transparency (commonly used in $^v$BN) parameters to explore the influence of marking parameters (intensity). The big size and low transparency represent the stronger backdoor-based watermark and marking effect.
As shown in Fig. \ref{size 5 trans 0.0} $\sim$ Fig. \ref{trans 0.6}, we compare the unlearning verification results when trigger size becomes smaller and trigger transparency becomes higher, i.e., the marking effect gets weaker,
the performance on the markers at the marking time  (in light blue) is relatively low, the performance change introduced by unlearning is smaller, thus the verified unlearning effect becomes weaker.
Therefore, we should choose some moderate marking parameters to enhance the marking effect, and further promote the unlearning verification credibility.

\begin{figure}[!htb] 
  \setlength{\abovecaptionskip}{-0.05cm}
\setlength{\belowcaptionskip}{-0.2cm} 
\vspace{-0.55cm} 
  \centering
  \subfigure[2 participants]{ 
      \label{client 2}
      \includegraphics[width=0.31\linewidth]{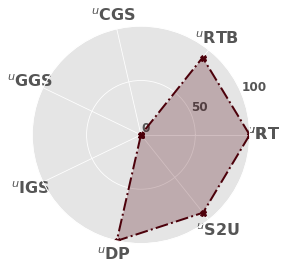}}
  \subfigure[10 participants]{
        \label{client 10}
        \includegraphics[width=0.31\linewidth]{./figure/radar/cifar_badnet_acc_radar.png}}
  \subfigure[20 participants]{
      \label{client 20}
      \includegraphics[width=0.31\linewidth]{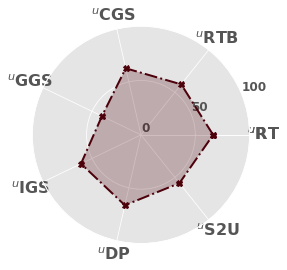}}
  \caption{Participant number influence on CIFAR10 dataset.}
  \label{Client number influence in BN}
\end{figure}

\noindent\textbf{Influence of the number of involved participants in FL: }
Fig. \ref{client 2} $\sim$ Fig. \ref{client 20} presents the unlearning verification results when the number of selected clients in each round is 2, 10, 20.
Through the comparison of the results, we can find that the performance difference on the $^v$BN markers becomes smaller with more participants involved in each round of FL.
The performance change caused by unlearning is more obvious when fewer participants upload their updates to the central server, as the result of average unlearning could not effectively erase so much memory (accounted for nearly half of the contribution to the global model when only 2 participants involved). Another reason is that the contribution of an individual is easily covered by others in large-system, i.e., other participants' updates would degrade the leaver's marking effect, and further reduce the performance difference caused by unlearning.

\subsection{Unlearning verification on leaving data with different levels of non-i.i.d. distribution}\label{Non-i.i.d. data }
\begin{figure}[!htb] 
  \setlength{\abovecaptionskip}{0.05cm}
  \centering
     \includegraphics[width=0.45\linewidth]{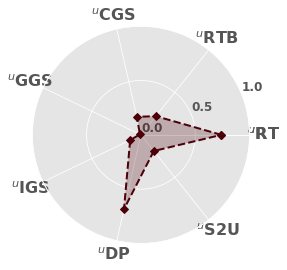}
      \includegraphics[width=0.45\linewidth]{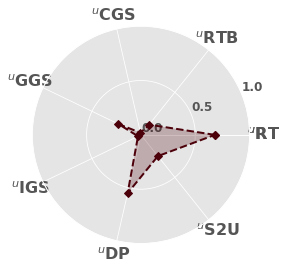}
  \caption{Unlearning verification on leaving data with different non-i.i.d. distribution extent. The left figures show the results on non-i.i.d data distribution with hyperparameter 0.9. The right figures show the results on non-i.i.d data distribution with hyperparameter 0.5 (more strict non-i.i.d.).  Both subfigures show the loss difference on the leaving data before and after unlearning by the corresponding unlearning method. }
  \label{non-i.i.d.}
\end{figure}

In our experiments, we directly use Dirichlet function \cite{minka2000estimating} with hyperparameter (0.9 and 0.5) to generate the individual data blocks satisfying different levels of non-i.i.d. distribution. The smaller the hyperparameter, the more strict the non-i.i.d. setting. As Fig. \ref{non-i.i.d.} shows, $^u$RT and $^u$DP maintain an obvious performance change on the leaving data,  $^u$GGS presents a bigger change under more strict non-i.i.d. distribution, the unlearning effect of other unlearning methods is still unobvious.
In a word, even though the leaving data (under a smaller hyperparameter in Dirichlet function) maintain larger different distribution with others' data, the unlearning effect solely on the leaving data is not so ideal, calling for more dedicated unlearning verification.

\section{Conclusion and Future Wok}

In this paper, we design and implement the first open-source platform --- \verifi, a unified federated unlearning and verification framework that allows systematic analysis of the verifiable amount of unlearning with different combinations of unlearning and verification methods. Based on \verifi, we conduct the first systematic study in the literature for verifiable federated unlearning, with 7 unlearning methods (including newly proposed $^u$S2U) and 5 verification methods (including newly proposed $^v$EM and $^v$FM),covering existing, adapted and newly proposed ones for both unlearning and verification.
Extensive experiments showed that our proposed $^u$S2U is an effective, efficient and secure federated unlearning method with little time cost, storage cost and negative impact on the original FL task.
The experiments also confirm the effectiveness of our proposed non-invasive $^v$EM methods for federated unlearning verification. 
The combination of $^v$EM and $^u$S2U yields so far the most promising approach for verifiable federated unlearning without tempering with the FL process, white-box model access or raising new security risks.

Research on verifiable federated unlearning is emerging and \verifi is able to serve as an open test bed for developing and benchmarking future federated unlearning and verification techniques. 
Following \verifi, there are a rich set of research opportunities to explore further, such as new unlearning and verification methods (concurrently taking the robustness, efficacy and additional fairness issue into consideration), free leaving of multiple participants, certification of federated unlearning, etc.

\clearpage
{\bibliographystyle{IEEEtranS}
\bibliography{sample}}

\begin{thebibliography}{10}
\providecommand{\url}[1]{#1}
\csname url@samestyle\endcsname
\providecommand{\newblock}{\relax}
\providecommand{\bibinfo}[2]{#2}
\providecommand{\BIBentrySTDinterwordspacing}{\spaceskip=0pt\relax}
\providecommand{\BIBentryALTinterwordstretchfactor}{4}
\providecommand{\BIBentryALTinterwordspacing}{\spaceskip=\fontdimen2\font plus
\BIBentryALTinterwordstretchfactor\fontdimen3\font minus
  \fontdimen4\font\relax}
\providecommand{\BIBforeignlanguage}[2]{{%
\expandafter\ifx\csname l@#1\endcsname\relax
\typeout{** WARNING: IEEEtranS.bst: No hyphenation pattern has been}%
\typeout{** loaded for the language `#1'. Using the pattern for}%
\typeout{** the default language instead.}%
\else
\language=\csname l@#1\endcsname
\fi
#2}}
\providecommand{\BIBdecl}{\relax}
\BIBdecl

\bibitem{abadi2016deep}
M.~Abadi, A.~Chu, I.~Goodfellow, H.~B. McMahan, I.~Mironov, K.~Talwar, and
  L.~Zhang, ``Deep learning with differential privacy,'' in \emph{ACM CCS},
  2016, pp. 308--318.

\bibitem{arpit2017closer}
D.~Arpit, S.~Jastrz{\k{e}}bski, N.~Ballas, D.~Krueger, E.~Bengio, M.~S. Kanwal,
  T.~Maharaj, A.~Fischer, A.~Courville, Y.~Bengio \emph{et~al.}, ``A closer
  look at memorization in deep networks,'' in \emph{ICML}.\hskip 1em plus 0.5em
  minus 0.4em\relax PMLR, 2017, pp. 233--242.

\bibitem{bagdasaryan2020backdoor}
E.~Bagdasaryan, A.~Veit, Y.~Hua, D.~Estrin, and V.~Shmatikov, ``How to backdoor
  federated learning,'' in \emph{International Conference on Artificial
  Intelligence and Statistics}.\hskip 1em plus 0.5em minus 0.4em\relax PMLR,
  2020, pp. 2938--2948.

\bibitem{blanchard2017machine}
P.~Blanchard, R.~Guerraoui, J.~Stainer \emph{et~al.}, ``Machine learning with
  adversaries: Byzantine tolerant gradient descent,'' in \emph{Advances in
  Neural Information Processing Systems}, 2017, pp. 119--129.

\bibitem{bonawitz2019towards}
K.~Bonawitz, H.~Eichner, W.~Grieskamp, D.~Huba, A.~Ingerman, V.~Ivanov,
  C.~Kiddon, J.~Kone{\v{c}}n{\`y}, S.~Mazzocchi, B.~McMahan \emph{et~al.},
  ``Towards federated learning at scale: System design,'' \emph{Proceedings of
  Machine Learning and Systems}, vol.~1, pp. 374--388, 2019.

\bibitem{bourtoule2021machine}
L.~Bourtoule, V.~Chandrasekaran, C.~A. Choquette-Choo, H.~Jia, A.~Travers,
  B.~Zhang, D.~Lie, and N.~Papernot, ``Machine unlearning,'' in \emph{2021 IEEE
  Symposium on Security and Privacy (SP)}.\hskip 1em plus 0.5em minus
  0.4em\relax IEEE, 2021, pp. 141--159.

\bibitem{brisimi2018federated}
T.~S. Brisimi, R.~Chen, T.~Mela, A.~Olshevsky, I.~C. Paschalidis, and W.~Shi,
  ``Federated learning of predictive models from federated electronic health
  records,'' \emph{International journal of medical informatics}, vol. 112, pp.
  59--67, 2018.

\bibitem{cao2018vggface2}
Q.~Cao, L.~Shen, W.~Xie, O.~M. Parkhi, and A.~Zisserman, ``Vggface2: A dataset
  for recognising faces across pose and age,'' in \emph{2018 13th IEEE
  international conference on automatic face \& gesture recognition (FG
  2018)}.\hskip 1em plus 0.5em minus 0.4em\relax IEEE, 2018, pp. 67--74.

\bibitem{cao2019ipguard}
X.~Cao, J.~Jia, and N.~Z. Gong, ``Ipguard: Protecting the intellectual property
  of deep neural networks via fingerprinting the classification boundary,''
  \emph{arXiv preprint arXiv:1910.12903}, 2019.

\bibitem{cao2015towards}
Y.~Cao and J.~Yang, ``Towards making systems forget with machine unlearning,''
  in \emph{2015 IEEE Symposium on Security and Privacy}.\hskip 1em plus 0.5em
  minus 0.4em\relax IEEE, 2015, pp. 463--480.

\bibitem{chen2021machine}
M.~Chen, Z.~Zhang, T.~Wang, M.~Backes, M.~Humbert, and Y.~Zhang, ``When machine
  unlearning jeopardizes privacy,'' in \emph{Proceedings of the 2021 ACM SIGSAC
  Conference on Computer and Communications Security}, 2021, pp. 896--911.

\bibitem{chen2017targeted}
X.~Chen, C.~Liu, B.~Li, K.~Lu, and D.~Song, ``Targeted backdoor attacks on deep
  learning systems using data poisoning,'' \emph{arXiv preprint
  arXiv:1712.05526}, 2017.

\bibitem{codella2018skin}
N.~C. Codella, D.~Gutman, M.~E. Celebi, B.~Helba, M.~A. Marchetti, S.~W. Dusza,
  A.~Kalloo, K.~Liopyris, N.~Mishra, H.~Kittler \emph{et~al.}, ``Skin lesion
  analysis toward melanoma detection: A challenge at the 2017 international
  symposium on biomedical imaging (isbi), hosted by the international skin
  imaging collaboration (isic),'' in \emph{2018 IEEE 15th international
  symposium on biomedical imaging (ISBI 2018)}.\hskip 1em plus 0.5em minus
  0.4em\relax IEEE, 2018, pp. 168--172.

\bibitem{combalia2019bcn20000}
M.~Combalia, N.~C. Codella, V.~Rotemberg, B.~Helba, V.~Vilaplana, O.~Reiter,
  C.~Carrera, A.~Barreiro, A.~C. Halpern, S.~Puig \emph{et~al.}, ``Bcn20000:
  Dermoscopic lesions in the wild,'' \emph{arXiv preprint arXiv:1908.02288},
  2019.

\bibitem{davis2017biology}
R.~L. Davis and Y.~Zhong, ``The biology of forgetting—a perspective,''
  \emph{Neuron}, vol.~95, no.~3, pp. 490--503, 2017.

\bibitem{deng2009imagenet}
J.~Deng, W.~Dong, R.~Socher, L.-J. Li, K.~Li, and L.~Fei-Fei, ``Imagenet: A
  large-scale hierarchical image database,'' in \emph{2009 IEEE conference on
  computer vision and pattern recognition}.\hskip 1em plus 0.5em minus
  0.4em\relax Ieee, 2009, pp. 248--255.

\bibitem{dong2020interactive}
E.~Dong, H.~Du, and L.~Gardner, ``An interactive web-based dashboard to track
  covid-19 in real time,'' \emph{The Lancet infectious diseases}, vol.~20,
  no.~5, pp. 533--534, 2020.

\bibitem{GDPR}
\BIBentryALTinterwordspacing
GDPR, ``Right to erasure (right to be forgotten),'' 2017. [Online]. Available:
  \url{https://gdpr-info.eu/art-17-gdpr/}
\BIBentrySTDinterwordspacing

\bibitem{ginart2019making}
A.~Ginart, M.~Guan, G.~Valiant, and J.~Y. Zou, ``Making ai forget you: Data
  deletion in machine learning,'' in \emph{Advances in Neural Information
  Processing Systems}, 2019, pp. 3518--3531.

\bibitem{goldberger2003efficient}
J.~Goldberger, S.~Gordon, H.~Greenspan \emph{et~al.}, ``An efficient image
  similarity measure based on approximations of kl-divergence between two
  gaussian mixtures.'' in \emph{ICCV}, vol.~3, 2003, pp. 487--493.

\bibitem{gu2017badnets}
T.~Gu, B.~Dolan-Gavitt, and S.~Garg, ``Badnets: Identifying vulnerabilities in
  the machine learning model supply chain,'' \emph{arXiv preprint
  arXiv:1708.06733}, 2017.

\bibitem{guerraoui2018hidden}
R.~Guerraoui, S.~Rouault \emph{et~al.}, ``The hidden vulnerability of
  distributed learning in byzantium,'' in \emph{International Conference on
  Machine Learning}.\hskip 1em plus 0.5em minus 0.4em\relax PMLR, 2018, pp.
  3521--3530.

\bibitem{harding2019understanding}
E.~L. Harding, J.~J. Vanto, R.~Clark, L.~Hannah~Ji, and S.~C. Ainsworth,
  ``Understanding the scope and impact of the california consumer privacy act
  of 2018,'' \emph{Journal of Data Protection \& Privacy}, vol.~2, no.~3, pp.
  234--253, 2019.

\bibitem{he2021deepobliviate}
Y.~He, G.~Meng, K.~Chen, J.~He, and X.~Hu, ``Deepobliviate: A powerful charm
  for erasing data residual memory in deep neural networks,'' \emph{arXiv
  preprint arXiv:2105.06209}, 2021.

\bibitem{kairouz2019advances}
P.~Kairouz, H.~B. McMahan, B.~Avent, A.~Bellet, M.~Bennis, A.~N. Bhagoji,
  K.~Bonawitz, Z.~Charles, G.~Cormode, R.~Cummings \emph{et~al.}, ``Advances
  and open problems in federated learning,'' \emph{arXiv preprint
  arXiv:1912.04977}, 2019.

\bibitem{kirkpatrick2017overcoming}
J.~Kirkpatrick, R.~Pascanu, N.~Rabinowitz, J.~Veness, G.~Desjardins, A.~A.
  Rusu, K.~Milan, J.~Quan, T.~Ramalho, A.~Grabska-Barwinska \emph{et~al.},
  ``Overcoming catastrophic forgetting in neural networks,'' \emph{Proceedings
  of the National Academy of Sciences}, vol. 114, no.~13, pp. 3521--3526, 2017.

\bibitem{koh2017understanding}
P.~W. Koh and P.~Liang, ``Understanding black-box predictions via influence
  functions,'' in \emph{International Conference on Machine Learning}.\hskip
  1em plus 0.5em minus 0.4em\relax PMLR, 2017, pp. 1885--1894.

\bibitem{krizhevsky2009learning}
A.~Krizhevsky, G.~Hinton \emph{et~al.}, ``Learning multiple layers of features
  from tiny images,'' 2009.

\bibitem{lecun1998gradient}
Y.~LeCun, L.~Bottou, Y.~Bengio, and P.~Haffner, ``Gradient-based learning
  applied to document recognition,'' \emph{Proceedings of the IEEE}, vol.~86,
  no.~11, pp. 2278--2324, 1998.

\bibitem{liu2021federaser}
G.~Liu, X.~Ma, Y.~Yang, C.~Wang, and J.~Liu, ``Federaser: Enabling efficient
  client-level data removal from federated learning models,'' in \emph{2021
  IEEE/ACM 29th International Symposium on Quality of Service (IWQOS)}.\hskip
  1em plus 0.5em minus 0.4em\relax IEEE, 2021, pp. 1--10.

\bibitem{liu2020federated}
G.~Liu, Y.~Yang, X.~Ma, C.~Wang, and J.~Liu, ``Federated unlearning,''
  \emph{arXiv preprint arXiv:2012.13891}, 2020.

\bibitem{liu2020learn}
Y.~Liu, Z.~Ma, X.~Liu, and J.~Ma, ``Learn to forget: User-level memorization
  elimination in federated learning,'' \emph{arXiv preprint arXiv:2003.10933},
  2020.

\bibitem{madry2017towards}
A.~Madry, A.~Makelov, L.~Schmidt, D.~Tsipras, and A.~Vladu, ``Towards deep
  learning models resistant to adversarial attacks,'' \emph{arXiv preprint
  arXiv:1706.06083}, 2017.

\bibitem{mcmahan2017communication}
B.~McMahan, E.~Moore, D.~Ramage, S.~Hampson, and B.~A. y~Arcas,
  ``Communication-efficient learning of deep networks from decentralized
  data,'' in \emph{Artificial Intelligence and Statistics}.\hskip 1em plus
  0.5em minus 0.4em\relax PMLR, 2017, pp. 1273--1282.

\bibitem{minka2000estimating}
T.~Minka, ``Estimating a dirichlet distribution,'' 2000.

\bibitem{parkhi2015deep}
O.~M. Parkhi, A.~Vedaldi, and A.~Zisserman, ``Deep face recognition,'' 2015.

\bibitem{regulation2018general}
P.~Regulation, ``General data protection regulation,'' \emph{Intouch}, 2018.

\bibitem{selvaraju2017grad}
R.~R. Selvaraju, M.~Cogswell, A.~Das, R.~Vedantam, D.~Parikh, and D.~Batra,
  ``Grad-cam: Visual explanations from deep networks via gradient-based
  localization,'' in \emph{Proceedings of the IEEE international conference on
  computer vision}, 2017, pp. 618--626.

\bibitem{shintre2019making}
S.~Shintre, K.~A. Roundy, and J.~Dhaliwal, ``Making machine learning forget,''
  in \emph{Annual Privacy Forum}.\hskip 1em plus 0.5em minus 0.4em\relax
  Springer, 2019, pp. 72--83.

\bibitem{shokri2017membership}
R.~Shokri, M.~Stronati, C.~Song, and V.~Shmatikov, ``Membership inference
  attacks against machine learning models,'' in \emph{2017 IEEE Symposium on
  Security and Privacy (SP)}.\hskip 1em plus 0.5em minus 0.4em\relax IEEE,
  2017, pp. 3--18.

\bibitem{sommer2020towards}
D.~M. Sommer, L.~Song, S.~Wagh, and P.~Mittal, ``Towards probabilistic
  verification of machine unlearning,'' \emph{arXiv preprint arXiv:2003.04247},
  2020.

\bibitem{sun2019can}
Z.~Sun, P.~Kairouz, A.~T. Suresh, and H.~B. McMahan, ``Can you really backdoor
  federated learning?'' \emph{arXiv preprint arXiv:1911.07963}, 2019.

\bibitem{toneva2018empirical}
M.~Toneva, A.~Sordoni, R.~T.~d. Combes, A.~Trischler, Y.~Bengio, and G.~J.
  Gordon, ``An empirical study of example forgetting during deep neural network
  learning,'' \emph{arXiv preprint arXiv:1812.05159}, 2018.

\bibitem{tschandl2018ham10000}
P.~Tschandl, C.~Rosendahl, and H.~Kittler, ``The ham10000 dataset, a large
  collection of multi-source dermatoscopic images of common pigmented skin
  lesions,'' \emph{Scientific data}, vol.~5, no.~1, pp. 1--9, 2018.

\bibitem{uchida2017embedding}
Y.~Uchida, Y.~Nagai, S.~Sakazawa, and S.~Satoh, ``Embedding watermarks into
  deep neural networks,'' in \emph{Proceedings of the 2017 ACM on International
  Conference on Multimedia Retrieval}, 2017, pp. 269--277.

\bibitem{wang2019interpret}
G.~Wang, ``Interpret federated learning with shapley values,'' \emph{arXiv
  preprint arXiv:1905.04519}, 2019.

\bibitem{wang2020attack}
H.~Wang, K.~Sreenivasan, S.~Rajput, H.~Vishwakarma, S.~Agarwal, J.-y. Sohn,
  K.~Lee, and D.~Papailiopoulos, ``Attack of the tails: Yes, you really can
  backdoor federated learning,'' \emph{Advances in Neural Information
  Processing Systems}, vol.~33, 2020.

\bibitem{Wang2020}
\BIBentryALTinterwordspacing
L.~Wang, Z.~Q. Lin, and A.~Wong, ``Covid-net: a tailored deep convolutional
  neural network design for detection of covid-19 cases from chest x-ray
  images,'' \emph{Scientific Reports}, vol.~10, no.~1, p. 19549, Nov 2020.
  [Online]. Available: \url{https://doi.org/10.1038/s41598-020-76550-z}
\BIBentrySTDinterwordspacing

\bibitem{warden2018speech}
P.~Warden, ``Speech commands: A dataset for limited-vocabulary speech
  recognition,'' \emph{arXiv preprint arXiv:1804.03209}, 2018.

\bibitem{WeBank}
\BIBentryALTinterwordspacing
WeBank, ``Webank and swiss re signed cooperation mou,'' 2019. [Online].
  Available:
  \url{https://finance.yahoo.com/news/webank-swiss-signed-cooperation-mou-112300218.html}
\BIBentrySTDinterwordspacing

\bibitem{xiao2017fashion}
H.~Xiao, K.~Rasul, and R.~Vollgraf, ``Fashion-mnist: a novel image dataset for
  benchmarking machine learning algorithms,'' \emph{arXiv preprint
  arXiv:1708.07747}, 2017.

\bibitem{xu2021federated}
J.~Xu, B.~S. Glicksberg, C.~Su, P.~Walker, J.~Bian, and F.~Wang, ``Federated
  learning for healthcare informatics,'' \emph{Journal of Healthcare
  Informatics Research}, vol.~5, no.~1, pp. 1--19, 2021.

\bibitem{yang2019federated}
Q.~Yang, Y.~Liu, T.~Chen, and Y.~Tong, ``Federated machine learning: Concept
  and applications,'' \emph{ACM Transactions on Intelligent Systems and
  Technology (TIST)}, vol.~10, no.~2, pp. 1--19, 2019.

\bibitem{yin2018byzantine}
D.~Yin, Y.~Chen, R.~Kannan, and P.~Bartlett, ``Byzantine-robust distributed
  learning: Towards optimal statistical rates,'' in \emph{International
  Conference on Machine Learning}.\hskip 1em plus 0.5em minus 0.4em\relax PMLR,
  2018, pp. 5650--5659.

\bibitem{zhang2017understanding}
C.~Zhang, S.~Bengio, M.~Hardt, B.~Recht, and O.~Vinyals, ``Understanding deep
  learning requires rethinking generalization,'' \emph{ICLR}, 2017.

\bibitem{zhang2018protecting}
J.~Zhang, Z.~Gu, J.~Jang, H.~Wu, M.~P. Stoecklin, H.~Huang, and I.~Molloy,
  ``Protecting intellectual property of deep neural networks with
  watermarking,'' in \emph{Proceedings of the 2018 on Asia Conference on
  Computer and Communications Security}, 2018, pp. 159--172.

\bibitem{zheng2001comparison}
F.~Zheng, G.~Zhang, and Z.~Song, ``Comparison of different implementations of
  mfcc,'' \emph{Journal of Computer science and Technology}, vol.~16, no.~6,
  pp. 582--589, 2001.

\end{thebibliography}

\section*{Appendices}

\subsection{Parameter}\label{parameter}

\textbf{MNIST} is a popular digit recognition dataset, adding up to 70000 32*32 gray images. Among them, 60000 pictures are used for training, and 10000 for testing \cite{xiao2017fashion}. 
\textbf{Cifar 10} is a widely recognized object classification dataset, consisting of 10 classes of 32*32 RGB pictures, 50000 for training and 10000 for testing  \cite{krizhevsky2009learning}. 
\textbf{ISIC} is composed of 4 classes skin images (Basal cell carcinoma, Melanoma, Benign keratosis-like keratosis and Melanocytic nevus), each class contains 2000 pictures, 1600 for training and 400 for testing \cite{tschandl2018ham10000, codella2018skin, combalia2019bcn20000}. 
\textbf{COVID} is the chest x-ray lung images, classified into 3 categories, including 1699 COVID-19, 6069 pneumonia, 8851 normal samples \cite{Wang2020}. 
We randomly sample 20 and 10 classes  from VGGFace and ImageNet to compose \textbf{VGGFace\_mini} and \textbf{ImageNet\_mini} \cite{cao2018vggface2, deng2009imagenet}. 
In \textbf{VGGFace\_mini}, there are totally 7023 224 *224 face images, 4916 for training, 2107 for testing.
\textbf{ImageNet\_mini} is composed by 13500 224 *224 RGB images, 13000 for training, 500 for testing.
\textbf{Speechcommand} is composed by 32 *32 spectrograms after MFCC \cite{zheng2001comparison}, 37005 for training, 9251 for testing.
To stimulate the non-i.i.d. data distribution, we also provide the Dirichlet distribution function \cite{minka2000estimating} to supply the unbalanced data to each participant.
The corresponding models are summarized in Table \ref{model architecture}.

Different unlearning methods need diverse parameters.
The $\left(\epsilon,\delta\right)$ parameter pair used in \emph{$^u$DP} is set $\left(0.2, 0\right)$ for MNIST, and $\left(0.1, 0\right)$ for other datasets.
$\lambda$ used in $^u${GS} is $0.01$ for all datasets.
The scaling ratio $\alpha$ and $\beta$ in $^u$S2U is set as $0.1$ and $1$ for all datasets.
Table \ref{Parameters of verification} summarizes the hyper-parameters used for different verification methods.

\begin{table}[!htb]
  \centering
  \caption{Parameters of verification}
  \label{Parameters of verification}
  \scalebox{0.6}{
  \begin{tabular}{c|c|c}
    \toprule
  \textbf{Verification}      & \textbf{Parameter}         & \textbf{Setting}                         \\ \midrule
                             & learning rate              & 0.01                                     \\
                             & mark epoch              & 100 for MNIST and CIFAR10 (50 for others)                                     \\
                             & optimizer                  & SGD                                      \\
                             & momentum                   & 0.9                                      \\
  \multirow{-5}{*}{Training} & weight decay               & 2.00E-04                                 \\ \midrule
                                 & embedding dim                  & 64                                       \\
                             & embedding layer                & features.conv1 (conv0 for ISIC)         \\
                             & projetcted matrix          & random                                   \\
  \multirow{-4}{*}{$^v$ME}       & penality loss ratio        & 0.05                                     \\ \midrule
                             & size                       & 5 for 32*32 (25 for 224*224) \\
                             & alpha                      & 0                \\
                             & toxic data percent         & 10\% for others and  30\%  for CIFAR10      \\
                             & backdoor learning rate  & 0.01 for others and  0.05  for CIFAR10 \\
  \multirow{-5}{*}{$^v$BN}       & target class               & 0                \\ \midrule
                            & the size of $\widetilde D$ & 400                                      \\
  \multirow{-2}{*}{$^v$SF}       & the size of $D_{\mathsf{a}}^{m}$                & 20        \\ \midrule
                  & number of boundary data & 100                                      \\
  \multirow{-2}{*}{$^v$BF}      &  loss gap $\gamma$            & 0.01                                     \\ \midrule
  
    $^v$FM                         & forgettable memory ratio       & 0.1                                      \\  \midrule
    \multirow{-1}{*}{$^v$EM}       & errorneous sample propportion $\kappa$           & 0.1                                      \\

  \bottomrule
                              
  \end{tabular}}
  \end{table}

\subsection{Federated Unlearning Verification by Comparing Parameter Differences}\label{sec:parameter_distance}
We compare the model parameter deviation before and after unlearning in Table \ref{The deviation of model parameters --- CIFAR10}. 
We could observe that except $^u$DP, $^u$RT and $^u$S2U, 
\emph{most unlearning methods have similar parameter deviation with $^u$NT and $^u$NF}, which means model parameters fail to reliably verify whether unlearning is effective. 
Thus, the unlearning effect cannot be directly observed from the model parameter deviation.

\begin{table*}[ht]
  \centering
  \caption{The model parameter difference after unlearning}
  \label{The deviation of model parameters --- CIFAR10}
  \scalebox{0.9}{
  \begin{tabular}{c|c|ccccccccc}
  \toprule
  \textbf{Dataset}  & \textbf{Method}  & \textbf{$^u$NF} & \textbf{$^u$NT} &\textbf{$^u$RT}  & \textbf{$^u$RTB}  &\textbf{$^u$CGS}  & \textbf{$^u$GGS}& \textbf{$^u$IGS} & \textbf{$^u$DP} & \textbf{$^u$S2U} \\ \midrule
  \multirow{2}{*}{CIFAR10} &Euclidean Distance     &1.44 & 1.55 & \cellcolor[HTML]{C0C0C0}29.37    & 1.48  & 1.52 & 1.47 & 1.49   & \cellcolor[HTML]{C0C0C0}334.33 & 3.47     \\
  &Cosine Similarity([0,1])     &0.99 & 0.99  & \cellcolor[HTML]{C0C0C0}0.60    & 0.99 &0.99 & 0.99  & 0.99 & \cellcolor[HTML]{C0C0C0}0.50&0.98\\ \midrule
  \multirow{2}{*}{SpeechCommand}  &Euclidean Distance &2.23 &2.29 &\cellcolor[HTML]{C0C0C0}5.09 & 2.00& 2.56& 2.23& 2.13&\cellcolor[HTML]{C0C0C0}34.90 & 2.40\\
  & Cosine Similarity([0,1]) &0.87 &0.92 &\cellcolor[HTML]{C0C0C0}0.76 &0.93 &0.88 &0.94 &0.94 &\cellcolor[HTML]{C0C0C0}0.38 & 0.95\\ \midrule
  \multirow{2}{*}{Covid}   &Euclidean Distance &1.49 &1.77 & \cellcolor[HTML]{C0C0C0}91.22&1.52 &2.30 & 2.03&1.51 &\cellcolor[HTML]{C0C0C0}334.28 & 5.54\\
  &Cosine Similarity([0,1])  &0.97 & 0.97&\cellcolor[HTML]{C0C0C0}0.53 & 0.97&0.96 & 0.96&0.98 & \cellcolor[HTML]{C0C0C0}0.47& 0.92\\ \midrule
  \multirow{2}{*}{VGGFace\_mini}&Euclidean Distance &3.39 &3.57 &\cellcolor[HTML]{C0C0C0}109.03 &3.34 & 3.49& 3.38& 3.36& \cellcolor[HTML]{C0C0C0}334.43& 13.95\\
  & Cosine Similarity([0,1]) &0.93 &0.93 &\cellcolor[HTML]{C0C0C0}0.35 & 0.93& 0.93& 0.93&0.93 & \cellcolor[HTML]{C0C0C0}0.44&0.86 \\ 
  \bottomrule
  \end{tabular}}
\end{table*}

\subsection{Federated Unlearning Verification by Membership Inference}\label{sec:membership_inference}
\begin{figure}[!htb]
\centering
\includegraphics[width=0.48\textwidth]{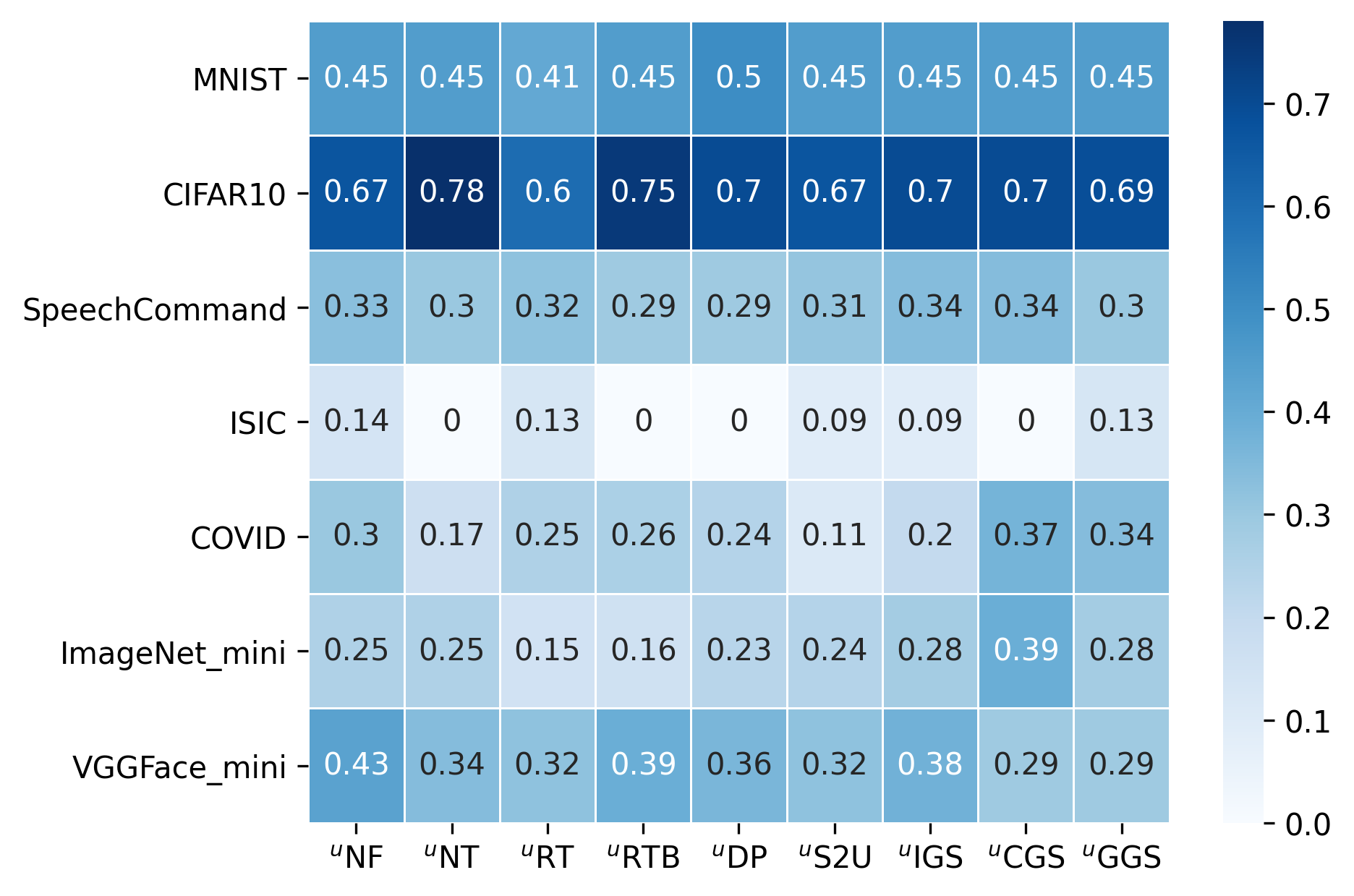}
\caption{Verifying unlearning from the perspective of privacy with each row representing one dataset, each column representing one unlearning method. The value represents the membership ratio between the deduced member data in the leaving data and all the leaving data.}
\label{member_privacy}
\end{figure}

We adopt membership inference in \cite{shokri2017membership} to verify unlearning from the perspective of privacy. In our experiments, the global model at the end of FL is treated as the shadow model. The training data and test data are directly regarded as the member and non-member data to simplify the process. As shown in Fig. \ref{member_privacy}, verifying federated unlearning from the perspective of privacy is infeasible. Since even after being unlearned by $^u$RT, which absolutely and completely removes the leaving data and retrains from scratch, the deduced membership ratio still would not drop significantly than $^u$NT, sometimes even higher. As for the reason, the similar data may belong to other participants in the federation, increasing the difficulty to verify unlearning in the aspect of privacy. It is worth mentioning that unlearning itself would cause the extra privacy concerns, which is out of scope of \verifi, we recommend reading the work \cite{chen2021machine}.

\subsection{Security risk of $^v$BN}\label{Security risk of $^v$BN}
\begin{figure}[!htb]
  \centering
  \includegraphics[width=0.48\textwidth]{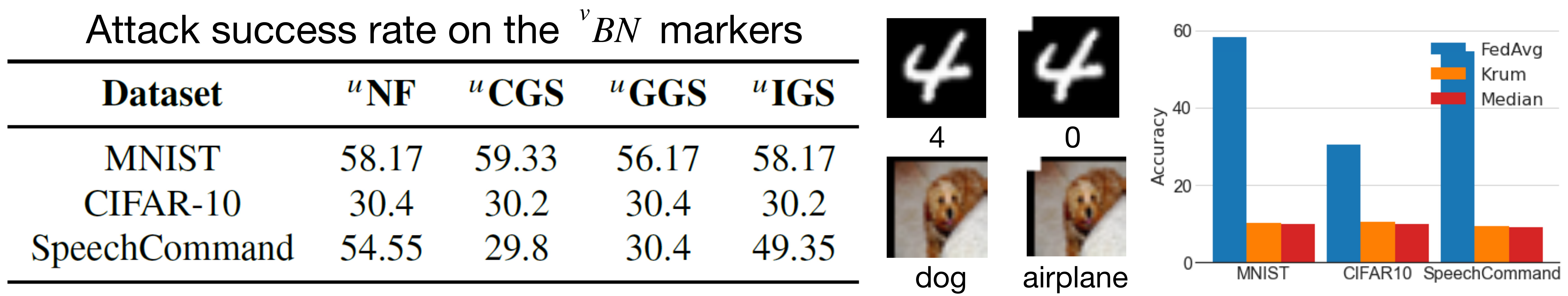}
  \caption{Security threat of $^v$BN}
  \label{invasive_security}
  \end{figure}

Apart from working as a watermark to verify unlearning, $^v$BN itself is a traditional backdoor attack widely studied in \cite{sun2019can,bagdasaryan2020backdoor,wang2020attack}.
Fig. \ref{invasive_security} shows that even at the end of FL, the backdoor (any sample patched with the trigger would be classified into the target class) still exists. 
Some unlearning methods cannot completely remove the security risk caused by the invasive marking method. Specifically, even with unlearning, the attack success rate of backdoor-based watermark even reaches 50\%. Fortunately, the security threat can be removed with the robust aggregation rules, such as Krum and Median, as shown in Fig. \ref{invasive_security}.

  \subsection{Verification effect difference between unique memory samples and leaving data}\label{Verification effect difference between unique memory samples and leaving data}
\begin{figure}[!htb] 
  \setlength{\belowcaptionskip}{-0.2cm} 
  \vspace{-0.15cm} 
    \centering
    \subfigure[$^v$EM]{ 
        \includegraphics[height=0.6in,width=0.2\linewidth]{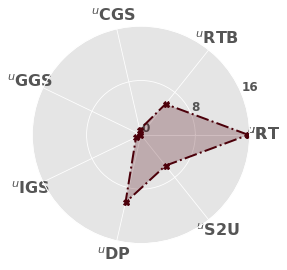}
        \includegraphics[height=0.6in,width=0.2\linewidth]{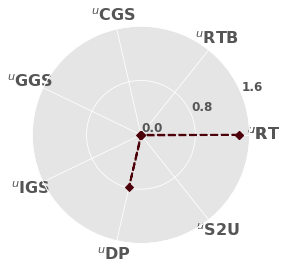}
        }
    \subfigure[$^v$FM]{
          \label{client 10}
          \includegraphics[height=0.6in,width=0.2\linewidth]{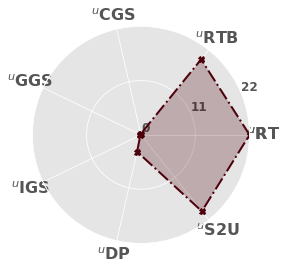}
          \includegraphics[height=0.6in,width=0.2\linewidth]{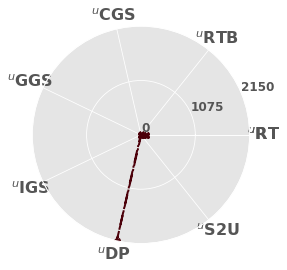}}
    \caption{Unlearning verification effect difference between leveraging the unique memory samples and leaving data on CIFAR10 dataset. (a) shows the unlearning verification effect of erroneous memory $^v$EM markers and leaving data, (b) shows the unlearning verification effect of forgettable memory $^v$FM markers and leaving data.}
    \label{unlearning verification effect difference}
  \end{figure}

  As shown in Section \ref{Proposed Unique Memory Markers.}, we have presented the details and causes of choosing the particular unique memory samples as the markers. Here, we discuss the concrete unlearning verification effect difference between leveraging the unique memory samples and the leaving data. Fig. \ref{unlearning verification effect difference} shows that the selected memory markers could maintain the better unlearning verification effect, since they can identify the unlearning performance of other unlearning methods besides $^u$RT and $^u$DP.
  
\subsection{Unlearning verification visualization}
\begin{figure*}[!htb] 
\vspace{-0.35cm}
   \setlength{\abovecaptionskip}{-0.0cm}
     \setlength{\belowcaptionskip}{-0.2cm} 
    \centering
    \subfigure[Example]{
        \label{example_horse}
        \includegraphics[width=0.08\linewidth]{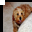}}
    \subfigure[Before Unlearning]{ 
        \label{Before Unlearning}
        \includegraphics[width=0.08\linewidth]{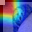}}
    \subfigure[$^u$NF]{
        \label{result_horse_none}
        \includegraphics[width=0.08\linewidth]{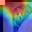}}
    \subfigure[$^u$RT]{
        \label{result_horse_dp}
        \includegraphics[width=0.08\linewidth]{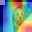}}
    \subfigure[$^u$RTB]{
        \label{result_horse_scale}
        \includegraphics[width=0.08\linewidth]{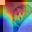}}
    \subfigure[$^u$CGS]{
        \label{result_horse_unlearning}
        \includegraphics[width=0.08\linewidth]{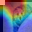}}
    \subfigure[$^u$GGS]{
        \label{result_horse_unlearning}
        \includegraphics[width=0.08\linewidth]{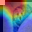}}
    \subfigure[$^u$IGS]{
      \label{result_horse_unlearning}
      \includegraphics[width=0.08\linewidth]{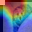}}
    \subfigure[$^u$DP]{
        \label{result_horse_unlearning}
        \includegraphics[width=0.08\linewidth]{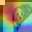}}
    \subfigure[$^u$S2U]{
        \label{result_horse_unlearning}
        \includegraphics[width=0.08\linewidth]{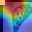}}
    \caption{Unlearning verification visualization in $^v$BN --- CIFAR10}
  
    \label{Unlearning verification visualization in $^v$BN --- CIFAR10}
  \end{figure*}

We show the unlearning verification effect via leveraging the interpretability technique --- Grad-CAM \cite{selvaraju2017grad} in Fig. \ref{Unlearning verification visualization in $^v$BN --- CIFAR10}.
These saliency maps of the selected forgotten individual sample, patched with a 5*5 white square trigger in $^v$BN, are computed based on the global model at the end of FL.
The original label of the sample is dog and the target class of the backdoor example is airplane.
Before unlearning, the memory about $\iota$ still exists as the backdoor sample is classified as the target class and the high attention area is mainly located on the trigger (see Fig. \ref{Before Unlearning}).
As Fig. \ref{Unlearning verification visualization in $^v$BN --- CIFAR10} shows, $^u$RT obtains the most explicit unlearning effect since the attention on the trigger, caused by the leaver $\iota$, totally disappears. $^u$RTB, $^u$DP and $^u$S2U apparently degrade the high attention on the trigger, not ideal like $^u$RT. 
The weakened attention can be owed to the gradually eliminated memory about the backdoor watermark introduced by $\iota$. The attention decrease on the trigger can also be observed in $^u$NF, $^u$CGS, $^u$GGS and $^u$IGS, however, not so obvious as other unlearning methods, as the result of unsatisfied unlearning.
Benefiting from the verification method, the unlearning effect can be explicitly presented with the visualization technique.

\subsection{Theoretical Explanation}\label{Theoretical Explanation}
We provide a theoretical unlearning verification explanation to help understand why unlearning could be verified on the $^v$BF markers or from the perspective of privacy.

\textbf{$^u$S2U and $^v$BF:}
We provide a simple theoretical explanation of how $^u$S2U quantitatively changes the decision boundary of the global model characterized by boundary examples ($^v$BF), which can then be verified with our proposed metrics. 

We assume that the central server applies FedAvg aggregation rule to update the global model of the next round: 
    \begin{equation}\label{eq:f2}
    \vw_{t+1} = \vw_t + \frac{1}{n}\sum_{i \in \left[ n \right]}(\vw_{t+1}^{(i)} - \vw_t)
    \end{equation}

$^u$S2U lowers the leaving participant $\mathsf{a}$'s contribution to the global model by reducing  $\mathsf{a}$'s local update:
\begin{shrinkeq}{-1ex}
\begin{equation}
\varphi \left( \vw_{t+1}^{(\mathsf{a})} - \vw_t \right) = \alpha\left( \vw_{t+1}^{(\mathsf{a})} - \vw_t \right) , \; t=t_{u},
\end{equation}
\end{shrinkeq}

\begin{shrinkeq}{-1ex}
\begin{equation}
\forall_{j \in \etC\diagup \mathsf{a}}  \left(\vw_{t+1}^{(j)} - \vw_t \right)= \beta\left(\vw_{T_{enabled}} - \vw_t\right), \; t=t_{u},
\end{equation}
\end{shrinkeq}
We set $\mathsf{a} = n$ and $\beta=1$ to facilitate understanding of the explanation.
$^v$BF characterizes the decision boundary of the local model $f^{(\mathsf{a})}$ using a subset of perturbed training samples close to the decision boundary. Arguably, the samples with relatively high and close top-2 class probabilities are boundary samples. $^v$BF generates the boundary markers satisfying:
\begin{equation}
    D^m_{\mathsf{a}} = \{(\vx, y)| (\vx, y) \in D_{\mathsf{a}}, |f^{(\mathsf{a})}_{top-1}(\vx + \sigma) - f^{(\mathsf{a})}_{top-2}(\vx + \sigma)| \leq \gamma \},
\end{equation}
where, $f^{(\mathsf{a})}_{top-1}(\vx + \sigma)$ and $f^{(\mathsf{a})}_{top-2}(\vx + \sigma)$ denote the top-1 and top-2 class probabilities respectively, $\sigma$ is the generated perturbation by PGD and $\gamma \in [0, 0.1)$ is a small positive value defining how close the two probabilities.
Since unlearning and verification are activated after $T_{enabled}$ in \verifi, when the global model has converged to a good solution, we can make a reasonable assumption that the boundary samples of  $\mathsf{a}$'s local model could also work as boundary samples to the converged global model $\vw_{t+1}$.

To simplify the complex derivation, we choose a binary classifier (assumed linearly around $\{ \vx_i,y_i \} \in D^m_{\mathsf{a}}, y_i \in \{ \pm 1\} $), $f(\vx_i) = \vw_t^T\vx_i+b$, 
\begin{equation}
y_i = \begin{cases}  +1,  &  \vw_t^T\vx_i+b \geq 1 \\  -1, &  \vw_t^T\vx_i+b \leq -1  \\ \end{cases}
\end{equation}
Then, we get $ y_i(\vw_t^T\vx_i+b) \geq 1$, the decision boundary distance between the two classes $\{+1, -1\}$ is $\frac{2}{\| \vw_t\|}$, the optimizer would optimize $\frac{2}{\| \vw_t\| }$ to enlarge the distance of decision boundary between two neighbor classes : $\max_{\vw_t} \{\frac{2}{\| \vw_t\|} \}$

\begin{figure}[!htb] 
  \centering
  \subfigure[Decision boundary before unlearning]{
      \label{Decison_boudary_before_unlearning}
      \includegraphics[width=0.45\linewidth]{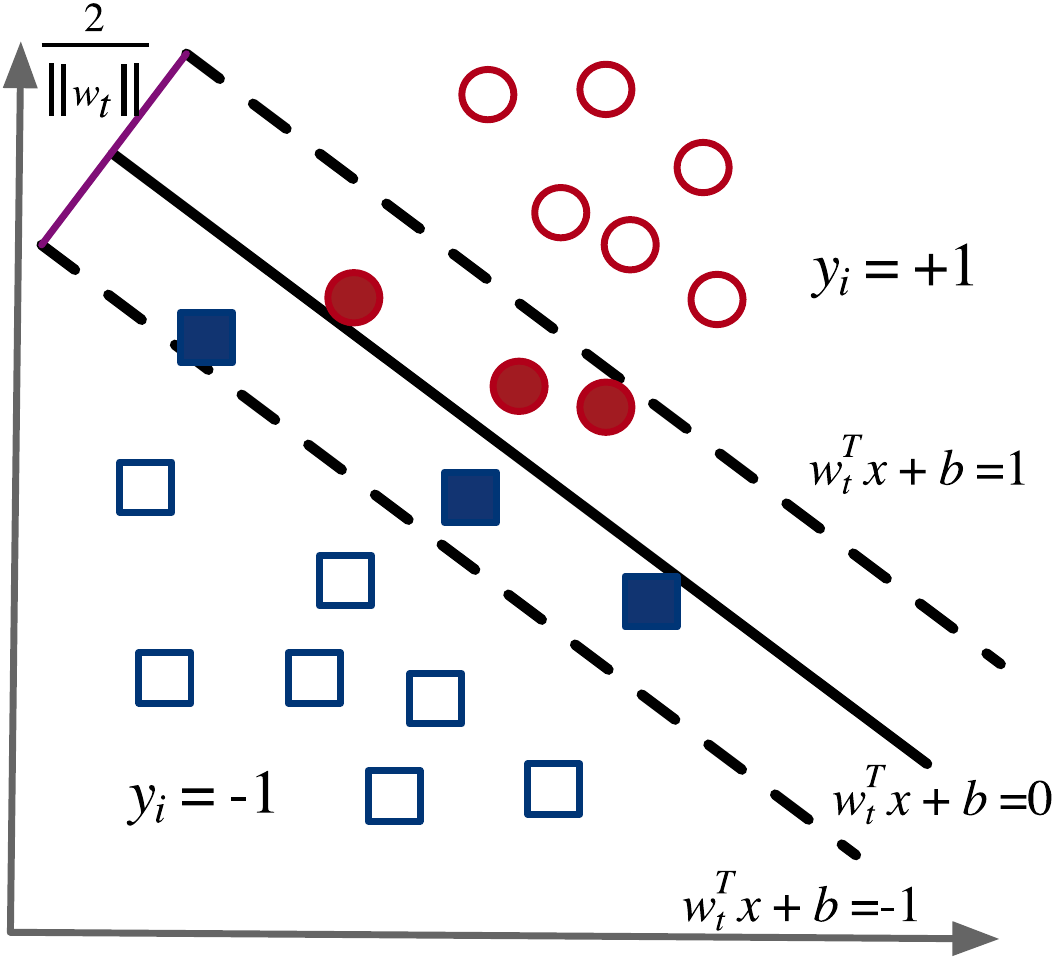}}
  \subfigure[Decision boundary after unlearning]{
      \label{Decison_boudary_after_unlearning}
      \includegraphics[width=0.45\linewidth]{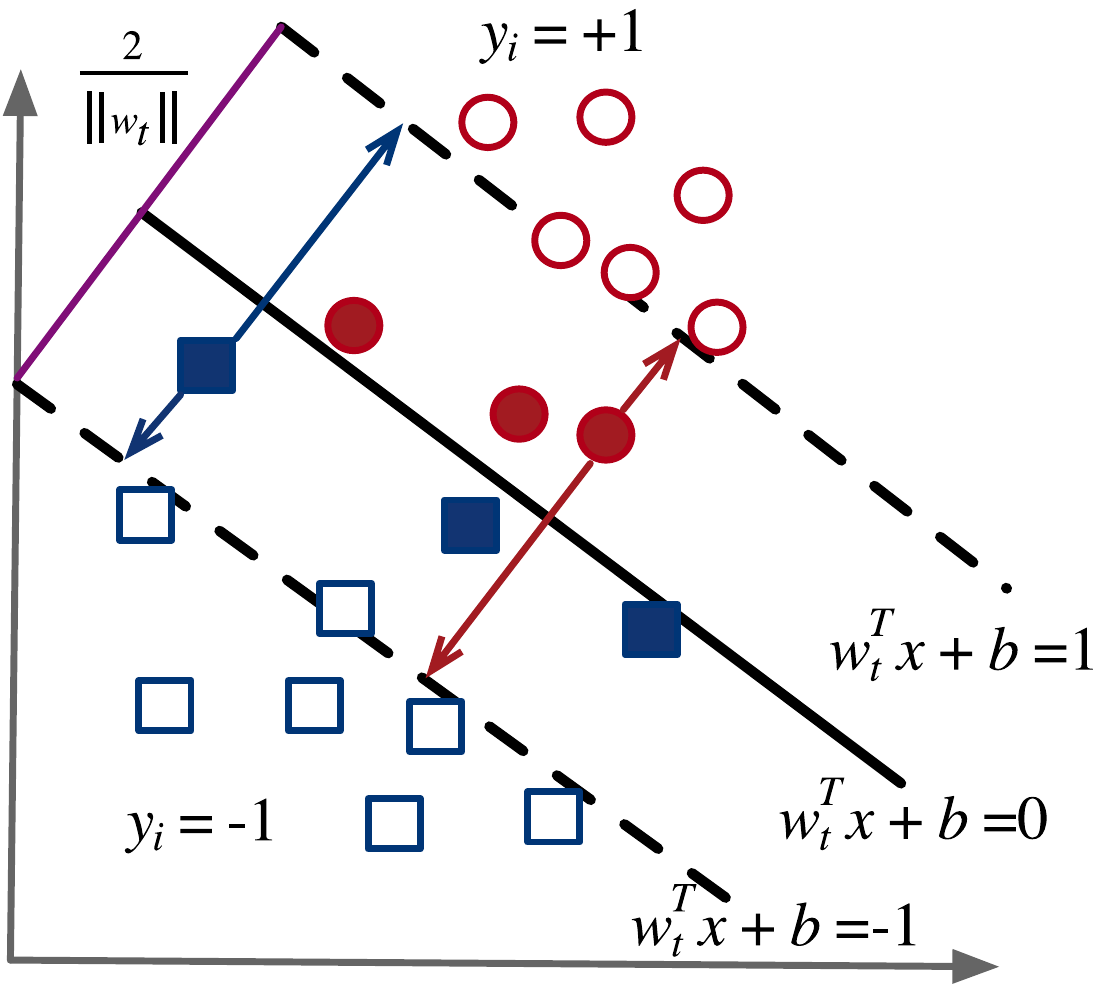}}

  \caption{Decision boundary change after unlearning}

  \label{Decison boudary change after unlearning}
\end{figure}

After launching $^u$S2U at $t$, the global model at $t+1$ is:
\begin{equation}
    \begin{aligned}
    \vw_{t+1} &= \vw_t + \frac{1}{n} \left(\sum_{i \in [n-1]}(\vw_{t+1}^{(i)} - \vw_t)  + \varphi \left( \vw_{t+1}^{(n)} - \vw_{t}  \right)  \right) \\
    &= \vw_t + \frac{ \sum\limits_{i \in [n]} (\vw_{t+1}^{(i)}-\vw_t)+ (\alpha-1)(\vw_{t+1}^{(n)} - \vw_t)  }{n} \\
    & \approx \vw_t+  \frac{ \sum\limits_{i \in [n-1]}(\vw_{t+1}^{(i)} - \vw_t) }{n} \\
    \end{aligned}
\end{equation}
The average local update from other $n-1$ participants is: $\overline{\vw_{t+1}^{[n-1]}} = \frac{ \sum_{i \in [n-1]} \vw_{t+1}^{(i)}  }{n-1} $, the global model update at $t+1$: $\vw_{t+1}-\vw_{t} = \frac{(n-1) \overline{\vw_{t+1}^{[n-1]}}}{n}$ $\textless$ $\overline{\vw_{t+1}^{[n-1]}}$, then $^u$S2U works by scaling down/up his own/others' update and further influences the global model update. The hyper-parameter $\alpha$ used in our experiment is $0.1$.
Since the global model $\vw_{t+1}$ is reduced after $^u$S2U, then the distance between the two classes of decision boundary $\frac{2}{\| \vw\| }$ is enlarged, as shown in Fig. \ref{Decison boudary change after unlearning}. Thus, the results on the original constructed boundary samples would change, we then focus on the result change to verify whether unlearning is successful.

\textbf{$^u$DP and privacy:}
We analyze the unlearning effect of $^u$DP from the privacy perspective, specifically, the information about whether the sample $\vz$ is in $\mathsf{a}$'s local training data is limited. Once $^u$DP works as an unlearning method, the upper threshold of membership information about $\vz$ is limited.
The data of $\mathsf{a}$ is $D_{\mathsf{a}} \sim \sD$, $\sD$ denotes distribution of $D_{\mathsf{a}}$, the inferer is $Inf$, $\vw_t$ is the global model, the model of $\mathsf{a}$ is $\vw_t^{\mathsf{a}} = \vw_t(D_{\mathsf{a}})$, $b$ is uniformly chosen from $\{0,1\}$ which denotes whether $\vz$ belongs to $\mathsf{a}$. $b = 0$ if $\vz \sim \sD$, $b = 1$ if $\vz \sim  D_{\mathsf{a}}$.
The membership inference result can be expressed as:
\begin{equation}
Mem(Inf, \vw_t,m, \sD)= \begin{cases} 1,  &  Inf(\vz,\vw_{\mathsf{a}},m, \sD) = b  \\  0, & Inf(\vz,\vw_{\mathsf{a}},m, \sD) \neq  b \\ \end{cases} 
\end{equation}
Then the membership advantage of $Inf$ can be expressed as the difference between $Inf$'s true and false positive rate:
\begin{equation}
Mem\_Adv(Inf,\vw_t,m, \sD) = Pr[Inf=0|b=0]-Pr[Inf=0|b=1]
\end{equation}
Then we give a short theoretical explanation of $^u$DP could impose a strict limit on the information about $\mathsf{a}$:
\begin{equation}
Mem\_Adv(Inf,\vw_t, m,\sD) \leq e^\epsilon -1
\end{equation}
Given $D_{\mathsf{a}} = (z_1, \cdots , z_m) $ and $\vz \sim D$, then $D_{\mathsf{a}}^{'} = (z_1, \cdots ,z_{i-1},\vz, z_{i+1}, \cdots, z_m) $, $\vw_t^{\mathsf{a} '} = \vw_t(D_{\mathsf{a}}^{'})$.
$Inf(\vz,\vw_t^{\mathsf{a}}, m, \sD)$ and $Inf(z_{i},\vw_t^{\mathsf{a} '}, m, \sD)$ have identical distributions for all $i \in [m]$, thus, 
\begin{equation}
Pr[Inf=0|b=0]=1-\sE[\frac{1}{m} \sum\limits_{i=1}^{m}Inf(z_{i},\vw_t^{\mathsf{a}}, m, \sD) ]
\end{equation}
\begin{equation}
  Pr[Inf=0|b=1]=1-\sE[\frac{1}{m} \sum\limits_{i=1}^{m}Inf(z_{i},\vw_t^{\mathsf{a} '}, m, \sD) ]
\end{equation}
Then,
\begin{equation}
Mem\_Adv  = \sE[\frac{1}{m} \sum\limits_{i=1}^{m} ( Inf(z_{i},\vw_t^{\mathsf{a} '}, m, \sD) - Inf(z_{i},\vw_t^{\mathsf{a}}, m, \sD) )]
\end{equation}
Assume that the local models of k participants in a federated learning round is ${\vw_t^{1}, \cdots ,\vw_t^{k}}$. $^u$DP ensures that for all $j \in [k]$, 
\begin{equation}
Pr[\vw_t^{\mathsf{a} '} = \vw_t^{j}] \leq e^{\epsilon}Pr[\vw_t^{\mathsf{a}} = \vw_t^{j}]
\end{equation}
Thus the membership advantage can be written as:
\begin{small}
\begin{equation}
  \begin{aligned}
&\sum\limits_{j=1}^{k} \sE[\frac{1}{m} \sum\limits_{i=1}^{m} (Pr[\vw_t^{\mathsf{a} '} = \vw_t^{j}] Inf(z_{i},\vw_{\mathsf{a}}^{'}, m, \sD) - Pr[\vw_t^{\mathsf{a}} = \vw_t^{j}]Inf(z_{i},\vw_t^{\mathsf{a}}, m, \sD) )] \\
&= \sum\limits_{j=1}^{k} \sE[\frac{1}{m} \sum\limits_{i=1}^{m} (Pr[\vw_t^{\mathsf{a} '} = \vw_t^{j}] Inf(z_{i},\vw_t^{j}, m, \sD) - Pr[\vw_t^{\mathsf{a}} = \vw_t^{j}]Inf(z_{i},\vw_t^{j}, m, \sD) )] \\
&= \sum\limits_{j=1}^{k} \sE[\frac{1}{m} \sum\limits_{i=1}^{m} (Pr[\vw_t^{\mathsf{a}'} = \vw_t^{j}]-Pr[\vw_t^{\mathsf{a}} = \vw_t^{j}]) Inf(z_{i},\vw_t^{j}, m, \sD) ] \\
&\leq \sum\limits_{j=1}^{k} \sE[\frac{1}{m} \sum\limits_{i=1}^{m} ( e^{\epsilon}Pr[\vw_t^{\mathsf{a}} = \vw_t^{j}]-Pr[\vw_t^{\mathsf{a}} = \vw_t^{j}]) Inf(z_{i},\vw_t^{j}, m, \sD) ] \\
&= \sum\limits_{j=1}^{k} \sE[\frac{1}{m} \sum\limits_{i=1}^{m}  (e^{\epsilon}-1)Pr[\vw_t^{\mathsf{a}} = \vw_t^{j}] Inf(z_{i},\vw_t^{j}, m, \sD) ] \\
\end{aligned}
\end{equation}
\end{small}
Then, $Mem\_Adv$ must be smaller than $e^{\epsilon}-1$ since $Inf(z_{i},\vw_t^{j}, m, \sD) \leq 1$, meaning the membership inference advantage is limited by the upper threshold. Thus, after unlearned by $^u$DP, the membership information about $\mathsf{a}$ is limited, unveiling the effectiveness of unlearning from the perspective of privacy.

\end{document}